\documentclass[iop]{emulateapj}

\usepackage{tikz}
\usepackage{natbib}
\usepackage{amsmath}
\usepackage{hyperref}
\usepackage{lineno}
\usepackage[percent]{overpic}
\usepackage{float}
\usepackage{wrapfig}

\usepackage[title]{appendix}

\newcommand{\Sect}[1]{Section~\ref{#1}}
\newcommand{\Eq}[1]{Equation~\ref{#1}}
\newcommand{\Fig}[1]{Figure~\ref{#1}}
\newcommand{\Tab}[1]{Table~\ref{#1}}

\newcommand{\project}[1]{\textsc{#1}}
\newcommand{\lsst}{\project{LSST}}

\newcommand{\des}{\project{DES}}
\newcommand{\Chippr}{\project{CHIPPR}}

\newcommand{\repo}[1]{\texttt{#1}}

\newcommand{\chippr}{\repo{chippr}}
\newcommand{\cosmolike}{\repo{CosmoLike}}
\newcommand{\emcee}{\repo{emcee}}

\newcommand{\python}{\textit{Python}}

\newcommand{\data}{\ensuremath{\vec{d}}}

\newcommand{\pr}[1]{\ensuremath{\mathrm{p}(#1)}}

\newcommand{\gvn}{\mid}
\newcommand{\integral}[2]{\ensuremath{\int #1 \mathrm{d} #2}}
\newcommand{\sz}{spec-$z$}

\newcommand{\pz}{photo-$z$}
\newcommand{\Pz}{Photo-$z$}
\newcommand{\zpdf}{\pz\ PDF}

\newcommand{\pzpdf}{\pz\ posterior PDF}
\newcommand{\Pzpdf}{\Pz\ posterior PDF}
\newcommand{\pzip}{\pz\ implicit posterior}
\newcommand{\nz}{$n(z)$}
\newcommand{\Nz}{$N(z)$}

\newcommand{\bvec}[1]{\ensuremath{\boldsymbol{#1}}}
\newcommand{\ndphi}{\bvec{\phi}}
\newcommand{\mmle}{marginalized maximum posterior estimate}

\begin{document}

\title{How to obtain the redshift distribution from probabilistic redshift estimates}

\author{Alex I. Malz\altaffilmark{1,2}}
\author{David W. Hogg\altaffilmark{2,3,4,5}}
\email{aimalz@astro.ruhr-uni-bochum.de}

\altaffiltext{1}{Ruhr-University Bochum, German Centre for Cosmological Lensing, Universit\"{a}tsstra{\ss}e 150, 44801 Bochum, Germany}
\altaffiltext{2}{Center for Cosmology and Particle Physics, Department of Physics, New York University, 726 Broadway, 9th floor, New York, NY 10003, USA}
\altaffiltext{3}{Simons Center for Computational Astrophysics, 162 Fifth Avenue, 7th floor, New York, NY 10010, USA}
\altaffiltext{4}{Center for Data Science, New York University, 60 Fifth Avenue, 7th floor, New York, NY 10003, USA}
\altaffiltext{5}{Max-Planck-Institut f\"ur Astronomie, K\"onigstuhl 17, D-69117 Heidelberg, Germany}

\begin{abstract}
A trustworthy estimate of the redshift distribution $n(z)$ is crucial for using weak gravitational lensing and large-scale structure of galaxy catalogs to study cosmology.
Spectroscopic redshifts for the dim and numerous galaxies of next-generation weak-lensing surveys are expected to be unavailable, making photometric redshift (photo-$z$) probability density functions (PDFs) the next-best alternative for comprehensively encapsulating the nontrivial systematics affecting photo-$z$ point estimation.
The established stacked estimator of $n(z)$ avoids reducing photo-$z$ PDFs to point estimates but yields a systematically biased estimate of $n(z)$ that worsens with decreasing signal-to-noise, the very regime where photo-$z$ PDFs are most necessary.
We introduce Cosmological Hierarchical Inference with Probabilistic Photometric Redshifts (\textsc{CHIPPR}), a statistically rigorous probabilistic graphical model of redshift-dependent photometry, which correctly propagates the redshift uncertainty information beyond the best-fit estimator of $n(z)$ produced by traditional procedures and is provably the only self-consistent way to recover $n(z)$ from photo-$z$ PDFs.
We present the \texttt{chippr} prototype code
, noting that the mathematically justifiable approach incurs computational expense.
The \textsc{CHIPPR} approach is applicable to any one-point statistic of any random variable, provided the prior probability density used to produce the posteriors is explicitly known; 
if the prior is implicit, as may be the case for popular photo-$z$ techniques, then the resulting posterior PDFs cannot be used for scientific inference.
We therefore recommend that the photo-$z$ community focus on developing methodologies that enable the recovery of photo-$z$ likelihoods with support over all redshifts, either directly or via a known prior probability density.
\end{abstract}


\maketitle


\section{Introduction}
\label{sec:intro}

Photometric redshift (\pz) estimation has been a staple of studies of galaxy evolution, large-scale structure, and cosmology since its conception half a century ago \citep{baum_photoelectric_1962}.  
An extremely coarse spectrum in the form of photometry in a handful of broadband filters can be an effective substitute for the time- and photon-intensive process of obtaining a spectroscopic redshift (\sz), a procedure that may only be applied to relatively bright galaxies.  
Once the photometric colors are calibrated against either a library of spectral energy distribution (SED) templates or a data set of spectra for galaxies with known redshifts, a correspondence between photometric colors and redshifts may be constructed, forming a trustworthy basis for \pz\ estimation or testing.

Calculations of correlation functions of cosmic shears and galaxy positions that constrain the cosmological parameters require large numbers of high-confidence redshifts of surveyed galaxies.  
Many more \pz s may be obtained in the time it would take to observe a smaller number of \sz s, and \pz s may be measured for galaxies too dim for accurate \sz\ confirmation, permitting the compilation of large catalogs of galaxies spanning a broad range of redshifts and luminosities.  
\Pz s have thus enabled the era of precision cosmology, heralded by weak gravitational lensing tomography and baryon acoustic oscillation peak measurements.  

\begin{figure}
	\begin{center}
		\includegraphics[width=0.45\textwidth]{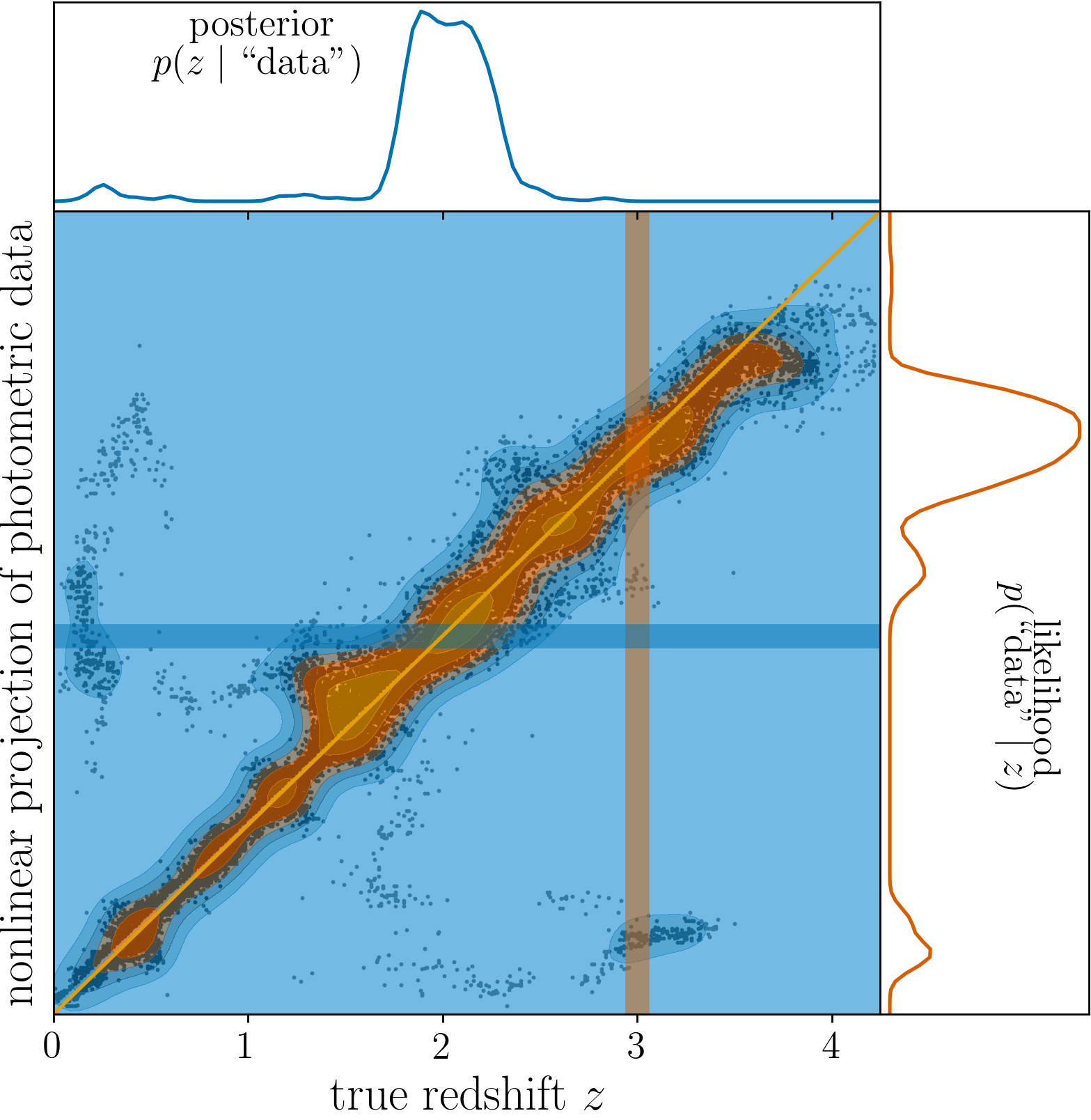}
		\caption{
			A generic probability space (darker in areas of higher probability density) of true redshift ($x$-axis) and a nonlinear projection of photometric data ($y$-axis), with vertical cuts and marginals (orange) indicating the construction of likelihoods and horizontal cuts and marginals (blue) indicating the construction of posteriors, with a theoretically perfect \pz\ estimate on the diagonal (yellow) for reference.
			The data points were extracted using WebPlotDigitizer \citep{rohatgi_webplotdigitizer_2019} from a \sz\ vs. \pz\ plot in \citet{jain_whole_2015}.
		}
		\label{fig:pedagogical_scatter}
	\end{center}
\end{figure}

However, \pz s are susceptible to inaccuracy and imprecision in the form of their inherent noisiness resulting from the decreasing signal-to-noise ratio with increasing redshift, coarseness of photometric filters, catastrophic errors in which galaxies of one SED at one redshift are mistaken for galaxies of another SED at a different redshift, and systematics introduced by observational techniques, data reduction processes, and training or template set limitations.  
Figure~\ref{fig:pedagogical_scatter} is an adaptation of the ubiquitous plots of \pz\ vs. \sz\ illustrating the assumptions underlying \pz\ estimation in general, that \sz s are a good approximation to true redshifts and \pz s represent special non-linear projections of observed photometry to a scalar variable that approximates the true redshift.

There are several varieties of generally non-Gaussian deviation from a trivial relationship between redshift and data in Figure~\ref{fig:pedagogical_scatter}, represented by a $y = x$ diagonal line.
The coarseness of the photometric filters causes scatter about the diagonal, with larger scatter perpendicular to the diagonal at redshifts where highly identifiable spectral features pass between the filters, as well as higher scatter at high redshifts where faint galaxies with large photometric errors are more abundant.
There are populations of outliers, far from the diagonal, comprised of galaxies for which the redshift estimate is catastrophically distinct from the true redshift, showing that outliers are not uniformly distributed nor restricted to long tails away from a Gaussian scatter.
And, though hardly perceptible in the plot, there is a systematic bias, wherein the average of the points would not lie on the diagonal but would be offset by a small bias, suggested by the trend of high-redshift points to lie below the diagonal.

Once propagated through the calculations of correlation functions of cosmic shear and galaxy positions, \pz\ errors are a dominant contributor to the total uncertainties reported on cosmological parameters \citep{abruzzo_impact_2019}.
As progress has been made on the influence of other sources of systematic error, the uncertainties associated with \pz s have come to dominate the error budget of cosmological parameter estimates made by current surveys such as \des\ \citep{hoyle_dark_2018}, \project{HSC} \citep{tanaka_photometric_2018}, and \project{KiDS} \citep{hildebrandt_kids-450:_2017}.
Based on the goals of a photometric galaxy survey, limits can be placed on the tolerance to these effects.
For example, the Science Requirements Document \citep{mandelbaum_weak_2017} states \lsst's requirements for the main cosmological sample, reproduced in Table~\ref{tab:lsstsrd}.

\begin{table}
	\begin{center}
		\caption{\Pz\ requirements for \lsst\ cosmology\\
				\citep{mandelbaum_weak_2017}.}
		\begin{tabular}{ll}
			Number of galaxies & $\approx 10^{7}$\\
			Root-mean-square error & $< 0.02 (1 + z)$\\
			$3 \sigma$ catastrophic outlier rate & $< 10\%$\\
			Canonical bias & $< 0.003 (1 + z)$\\
		\end{tabular}
		\label{tab:lsstsrd}
	\end{center}
\end{table}

Much effort has been dedicated to improving \pz s, though they are still most commonly obtained by a maximum likelihood estimator (MLE) based on libraries of galaxy SED templates, with conservative approaches to error estimation.
The presence of galaxies whose SEDs are not represented by the template library tends to lead to catastrophic outliers distributed like the horizontally oriented population of \Fig{fig:pedagogical_scatter}.
For data-driven approaches, training sets that are incomplete in redshift coverage tend to result in catastrophic outliers like the vertically oriented population of \Fig{fig:pedagogical_scatter}.
The approaches of using a training set versus a template library are related to one another by \citet{budavari_unified_2009}.
Sophisticated Bayesian techniques and machine learning methods have been employed to improve precision \citep{carliles_random_2010} and accuracy \citep{sadeh_annz2:_2016}, while other advances have focused on identifying and removing catastrophic outliers when using \pz s for inference \citep{gorecki_new_2014}. 

The probability density function (PDF) in redshift space for each galaxy, commonly written as $\pr{z}$, is an alternative to the MLE (with or without presumed Gaussian error bars) \citep{koo_photometric_1999}.
This option is favorable because it contains more potentially useful information about the uncertainty on each galaxy's redshift, incorporating our understanding of precision, accuracy, and systematic error.
However, denoting \zpdf s as ``$\pr{z}$'' is an abuse of notation, as it does not adequately convey what information is being used to constrain the redshift $z$; 
\zpdf s are \textit{posterior} PDFs, conditioned on the photometric data and prior knowledge.
In terms of \Fig{fig:pedagogical_scatter}, \zpdf s are horizontal cuts, probabilities of redshift conditioned on a specific value of data, i.e. posteriors $\pr{z \gvn \data}$, which constrain redshifts, whereas vertical cuts through this space are probabilities of data conditioned on a specific redshift, i.e. likelihoods $\pr{\data \gvn z}$, from which photometric data is actually drawn.

\Pzpdf s have been produced by completed surveys \citep{hildebrandt_cfhtlens:_2012, sheldon_photometric_2012} and will be produced by ongoing and upcoming surveys \citep{abell_lsst_2009, carrasco_kind_exhausting_2014, bonnett_redshift_2016, masters_mapping_2015}.  
\Pzpdf s are not without their own shortcomings, however, including the resources necessary to calculate and record them for large galaxy surveys \citep{carrasco_kind_sparse_2014, malz_approximating_2018} and the divergent results of each method used to derive them \citep{hildebrandt_phat:_2010, dahlen_critical_2013, sanchez_clustering_2013, bonnett_redshift_2016, tanaka_photometric_2018}.  
Though the matter is outside the scope of this paper, reviews of various methods have been presented in the literature \citep{sheldon_photometric_2012, ball_robust_2008, carrasco_kind_tpz:_2013, carrasco_kind_exhausting_2014, schmidt_evaluation_2020}.
The most concerning weakness of \pzpdf s, however, is their usage in the literature, which is at best inconsistent and at worst incorrect.  

Though their potential to improve estimates of physical parameters is tremendous, \pzpdf s have been applied only to a limited extent, most often by reduction to familiar point estimates.
If the true redshifts $\{z_{j}^{\dagger}\}$ of galaxies $j$ are known, then their redshift PDFs are well-approximated by delta functions $\{\delta(z, z_{j}^{\dagger})\}$ centered at the true redshift\footnote{Note that \sz s are not the same as known true redshifts; the PDFs of \sz s would be narrow and almost always unimodal, but they would not be delta functions due to observational errors.}, and the redshift distribution is effectively approximated by a histogram or other interpolation of the delta functions $\{\delta(z, z_{j}^{\dagger})\}$.
When \pzpdf s are available instead of true redshifts, the simplest approach reduces them to point estimates $\{\hat{z}_{i}\}$ of redshift by using $\delta(z, \hat{z}_{j})$ in place of $\delta(z, z_{j}^{\dagger})$.
Though it is most common for $\hat{z}_{j}$ to be the maximum or \textit{mode} of the \pzpdf, there are other, more principled point estimate reduction procedures \citep{tanaka_photometric_2018}.

Regardless of how it is done, any procedure that reduces \pzpdf s to point estimates discards valuable information about the uncertainty on redshift.
\Pzpdf s have also been used to form selection criteria of samples from galaxy surveys without propagation through the calculations of physical parameters \citep{van_breukelen_reliable_2009, viironen_high_2015}.  
Probability cuts on Bayesian quantities are not uncommon \citep{leung_bayesian_2017, dipompeo_quasar_2015}, but that procedure does not fully take advantage of all information contained in a probability distribution for parameter inference.

The most prevalent application of \pzpdf s that preserves their information content is the estimation of the \textit{redshift distribution function \Nz}, or, interchangably, its normalized cousin the \textit{redshift density function \nz}.
\nz\ is used to calculate the redshift calibration bias $b_{z}$ between the true and observed critical surface densities in galaxy-galaxy lensing \citep{mandelbaum_precision_2008} and the geometric lens efficiency $g_{k}(\chi)$ in tomographic weak lensing by large-scale structure \citep{benjamin_cfhtlens_2013}.
\Nz\ may be used to validate survey selection functions used in generation of realistic, multi-purpose mock catalogs \citep{norberg_2df_2002}.
As a key input to the traditional calculation of the power spectra of weak gravitational lensing and large-scale structure, the accuracy and precision to which \Nz\ is estimated can strongly impact our constraints on the parameters of cosmological models \citep{bonnett_using_2015,  masters_mapping_2015, viironen_high_2015, asorey_galaxy_2016, bonnett_redshift_2016, yang_calibrating_2018}, so it is unsurprising that this last application dominates the canonical bias requirement of Table~\ref{tab:lsstsrd}.
Even with \pz s adhering to the \lsst\ requirements of \Tab{tab:lsstsrd}, the degree to which constraints on the cosmological parameters can advance is limited by the accuracy and precision to which \nz\ is known \citep{abruzzo_impact_2019}.

Though it is traditional to estimate \nz\ from \pz\ point estimates \citep{abruzzo_impact_2019}, it has become more common to use \pzpdf s directly to calculate the conceptually simple but mathematically inconsistent \citep{hogg_data_2012} \textit{stacked estimator} $\hat{n}(z)$ of the redshift density function \citep{lima_estimating_2008}
\begin{align}
\label{eqn:stack}
\hat{n}(z) &= \frac{1}{J} \sum_{j = 0}^{J} \pr{z}_{j}
\end{align}
for a sample of $J$ galaxies $j$, or, equivalently, the redshift distribution function $\hat{N}(z) = J \hat{n}(z)$, by effectively averaging the \pzpdf s.
This summation procedure has been used extensively in cosmological analyses with photometric galaxy samples \citep{mandelbaum_precision_2008, benjamin_cfhtlens_2013, kelly_weighing_2014}.

Despite the growing prevalence of \pzpdf\ production, no implementation of inference using \pzpdf s has yet been presented with a mathematically consistent methodology.  
This paper challenges the logically invalid yet pervasive analysis procedure of stacking \pzpdf s by presenting and validating a hierarchical Bayesian technique for the use of \pzpdf s\ in the inference of \nz, yielding a method applicable to arbitrary one-point statistics relevant to cosmology, large-scale structure, and galaxy evolution; future work will extend this methodology to higher-order statistics.
We aim to develop a clear methodology guiding the use of \pzpdf s in inference so they may be utilized effectively by the cosmology community.
Though others have approached the problem before \citep{leistedt_hierarchical_2016, leistedt_hierarchical_2019}, the method presented here differs in that it makes use of any existing catalog of \pzpdf s, rather than requiring a simultaneous derivation of the \pzpdf s and the redshift distribution, making it preferable to ongoing surveys for which there may be inertia preventing a complete restructuring of the analysis pipeline.

In Section~\ref{sec:meth}, we present the \Chippr\ model for characterizing the full posterior probability landscape of \Nz\ using \pzpdf s. 
In Section~\ref{sec:application}, we present the \chippr\ implementation of the \Chippr\ model and the experimental set up by which we validate it, including the forward modeling of mock \pzpdf s.
In Section~\ref{sec:alldata}, we present a number of informative test cases and compare the results of \chippr\ with alternative approaches.
In Section~\ref{sec:results}, we stress-test the \Chippr\ model under nontraditional conditions.
Finally, in Section~\ref{sec:con}, we make recommendations for future research involving \nz\ estimation.

\section{Model}
\label{sec:meth}

Consider a survey of $J$ galaxies $j$, each with photometric data $\data_{j}$; 
thus the entire survey over some solid angle produces the ensemble of photometric magnitudes (or colors) and their associated observational errors $\{\data_{j}\}$.  
Each galaxy $j$ has a redshift parameter $z_{j}$ that we would like to learn.  
The distribution of the ensemble of redshift parameters $\{z_{j}\}$ may be described by the hyperparameters defining the redshift distribution function \nz\ that we would like to quantify.
The redshift distribution function \nz\ is the number of galaxies per unit redshift, effectively defining the evolution in the number of galaxies convolved with the selection function of the sample \citep{menard_clustering-based_2013}.  

In \Sect{sec:forward}, we establish a forward model encapsulating the causal relationship between \nz\ and photometry $\data$.  
In \Sect{sec:prob}, we present the directed acyclic graph of this probabilistic generative model and interpret the corresponding mathematical expression, whose full derivation may be found in the Appendix.
In \Sect{sec:limitations}, we summarize the necessary assumptions of the model.

\subsection{Forward Model}
\label{sec:forward}

We begin by reframing the redshift distribution \nz\ from a probabilistic perspective.
Here we define a redshift density \nz\ as the normalized probability density
\begin{equation}
\label{eqn:nz}
\int_{-\infty}^{\infty}\ n(z)\ dz\ \equiv\ \frac{1}{J}\ \int_{-\infty}^{\infty}\ \sum_{j=1}^{J}\ \delta(z_{j},\ z)\ dz = 1
\end{equation}
of finding a galaxy $j$ in a catalog of $J$ galaxies having a redshift $z$.
We believe that galaxy redshifts are indeed sampled, or drawn, from \nz, making it a probability density over redshift; this fact can also be confirmed by dimensional analysis of \Eq{eqn:nz}, as suggested in \citet{hogg_data_2012}.

We may without loss of generality impose a parameterization
\begin{equation}
\label{eqn:fz}
f(z; \ndphi)\ \equiv\ n(z)
\end{equation}
in terms of some parameter vector $\ndphi$.
At this point, the parameter vector is quite general and may represent coefficients in a high-order polynomial as a function of redshift, a set of means and variances defining Gaussians that sum to the desired distribution, a set of histogram heights that describe a binned version of the redshift distribution function, etc.
Upon doing so, we may rewrite \Eq{eqn:fz} as 
\begin{equation}
\label{eqn:pz}
z_{j}\ \sim\ \pr{z \gvn \ndphi}\ \equiv\ f(z; \ndphi),
\end{equation}
a probability density over redshift conditioned on the parameters $\ndphi$ specifying \nz.
Note that $z_{j}$ does not depend on the redshift $z_{j'}$ of some other galaxy $j' \neq j$, a statement of the causal independence of galaxy redshifts from one another.

In addition to believing \nz\ is a PDF from which redshifts are drawn, we also believe that there is some higher dimensional probability space $\pr{z, \data}$ of redshift $z$ and photometric data vectors $\data$, which may be any combination of fluxes, magnitudes, colors, and their observational errors.
Under this framework, \nz\ is equivalent to an integral
\begin{equation}
\label{eqn:integral}
n(z)\ =\ \integral{\pr{z, \data}}{\data}
\end{equation}
over the dimension of data in that joint probability space.
Note that galaxies may have different observational data despite sharing the same redshift, and that galaxies at different redshifts may have identical photometry; 
the space $\pr{z, \data}$ need not be one-to-one.
We assume a stronger version of statistical independence here, that draws $(z_{j}, \data_{j})$ are independent of draws $(z_{j'}, \data_{j'})$ in this space; 
the data and redshift of each galaxy are independent of those of other galaxies.

However, this problem has additional causal structure that we can acknowledge.
The photometry results from the redshifts, not the other way around.
This is the fundamental assumption upon which \pz\ estimation is based.
The forward model corresponds to first drawing redshifts according to \Eq{eqn:pz} and then drawing data from the likelihood
\begin{equation}
\label{eqn:pzpdf}
\data_{j}\ \sim\ \pr{\data \gvn z_{j}}
\end{equation}
of photometry conditioned on redshift, illustrated in Figure~\ref{fig:pedagogical_scatter}.

This description of the physical system corresponds to a forward model by which we actually believe photometry is generated:
\begin{enumerate}
	\item There exists a redshift distribution \nz\ with parameters $\ndphi$.
	\item Galaxy redshifts $\{z_{j}\}$ are independent draws from $\pr{z \gvn \ndphi}$.
	\item Galaxy photometry $\data_{j}$ is drawn from the likelihoods $\pr{\data_{j} \gvn z}$.
\end{enumerate}

\subsection{Probabilistic Model}
\label{sec:prob}

\begin{figure}
	\begin{center}
		\includegraphics[height=0.25\textheight]{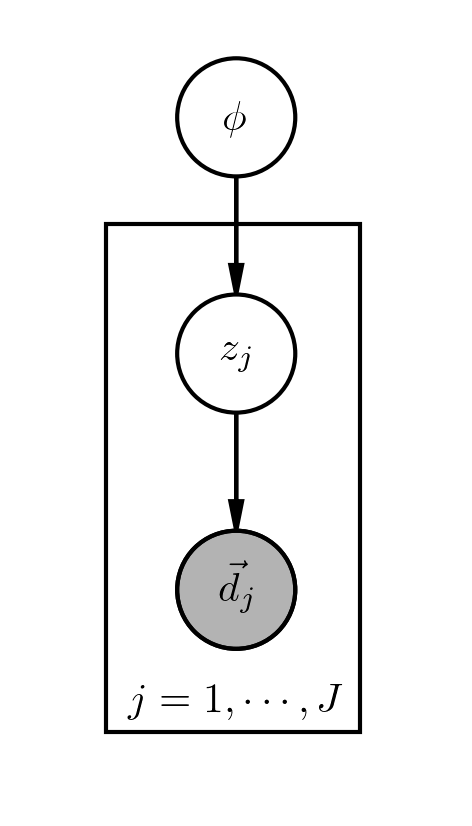}
		\caption{The directed acyclic graph of the CHIPPR model, where circles indicate random variables and arrows indicate causal relationships.
			The redshift distribution \nz\ parameterized by hyperparameters $\ndphi$ exists independent of the survey of $J$ galaxies, indicated as a box.  
			The redshifts $\{z_{j}\}$ of all galaxies in the survey are latent variables independently drawn from the redshift distribution, which is a function of $\ndphi$. 
			The photometric data $\data_{j}$ for each galaxy is drawn from a function of its redshift $z_{j}$ and observed, indicated by a shaded circle.
			}
		\label{fig:pgm}
	\end{center}
\end{figure}

A forward model such as that of \Sect{sec:forward} corresponds to a probabilistic graphical model (PGM), represented by a directed acyclic graph (DAG) as in \Fig{fig:pgm}.
A DAG conveys the causal relationships between physical parameters and, like a Feynman diagram in the context of particle physics, is a shorthand for mathematical relationships between variables.
The photometric data $\data_{j}$ of a galaxy is drawn from some function of its redshift $z_{j}$, independent of other galaxies' data and redshift.
Both data and redshift are random variables, but data is the one that we observe and redshift is not directly observable.
In this problem, we don't care about further constraining the redshifts of individual galaxies, only the redshift distribution \nz, so we consider redshift to be a \textit{latent variable}.
Because the parameters $\ndphi$ that we seek are causally separated from the data by the latent variable of redshift, we call them \textit{hyperparameters}.

The problem facing cosmologists is to determine the true value of $\ndphi$ from observing the photometry $\{\data_{j}\}$ of a large sample of $J$ galaxies $j$.
To self-consistently propagate the uncertainty in the inference of redshift, however, it is more appropriate to estimate the posterior $\pr{\ndphi \gvn \{\data_{j}\}}$ over all possible values of $\ndphi$ conditioned on all the observed data $\{\data_{j}\}$ available in a generic catalog.
In order to use the DAG of \Fig{fig:pgm} to derive an expression for $\pr{\ndphi \gvn \{\data_{j}\}}$ in terms of \pzpdf s, we must introduce two more concepts, confusingly named the \textit{implicit prior} and the \textit{prior probability density} (\textit{prior PDF}), elaborated upon below.

When we constrain the redshift of a galaxy using its observed photometric data $\data_{j}$, we are effectively estimating a posterior $\pr{z \gvn \data_{j}}$, the probability of an unknown quantity conditioned on the quantity we have in hand, i.e the photometric data.
This posterior is effectively a marginalization with respect to redshift at a given value of $\data = \data_{j}$ of the \textit{empirical frequency distribution} $\pr{z, \data \gvn \ndphi^{\dagger}}$, the joint probability density corresponding to the true redshift distribution parameterized by $\ndphi^{\dagger}$, which exists in nature but need not be known.

As the hyperparameters $\ndphi^{\dagger}$ of the true redshift distribution are in general unknown, the investigator seeking to estimate a posterior $\pr{z \gvn \data_{j}}$ must have a model $\phi^{*}$ for the general relationship between redshifts and photometry, whether empirical, as is the case for machine learning \pzpdf\ methods, or analytic, as is the case for template-based \pzpdf\ methods.
If we were to marginalize over the photometry in $\pr{\data, z}$, we would obtain a one-dimensional PDF $\pr{z \gvn \ndphi^{*}}$ over redshift, which can by definition be parameterized by the same functional form as \nz, for some $\ndphi^{*}$ specific to the estimation procedure that may or may not bear any relation to the hyperparameters $\ndphi^{\dagger}$ of the true \nz.
Rather, $\ndphi^{*}$ is a consequence of the generative model for how photometry results from redshift, including the influence of intrinsic galaxy spectra and instrumental effects. 

We call $\pr{z \gvn \ndphi^{*}}$ the \textit{implicit prior}, as it is rarely explicitly known nor chosen by the researcher\footnote{For template-based methods, the implicit prior is often an explicitly known input to the algorithm, engineered as an initial guess for the true $\ndphi$, with an aim for a realistic choice guided by an earlier spectroscopic survey.  
(See \citet{benitez_bayesian_2000} for more detail.)
It may thus be more appropriate to call it an \textit{interim prior}, but we will use the former term throughout this paper for generality.}
Because the implicit prior is unavoidable and almost inherently not uninformative, the \pzpdf s reported by any method must be \textit{implicit posteriors} ${\pr{z \gvn \data, \ndphi^{*}}}$ weighted by the implicit prior.

The prior probability density $\pr{\ndphi}$ is a more familiar concept in astronomy; to progress, we will have to choose a prior probability density over all possible values of the hyperparameters $\ndphi$.
This prior need not be excessively proscriptive; for example, it may be chosen to enforce smoothness at physically motivated scales in redshift without imposing any particular region as over- or under-dense.

With inputs of the \pzip\ catalog $\{\pr{z \gvn \data, \ndphi^{*}}\}$, the implicit prior $\pr{z \gvn \ndphi^{*}}$, and the prior PDF $\pr{\ndphi}$, we thus aim to obtain the posterior probability $\pr{\ndphi \gvn \{\data_{j}\}}$ of the redshift density function given all the photometric data.
By performing the derivation of the Appendix, we arrive at the desired expression
\begin{equation}
\label{eqn:fullpost}
\pr{\ndphi \gvn \{\data_{j}\}} \propto \pr{\ndphi} \integral{\prod_{j=1}^{J} \frac{\pr{z \gvn \data_{j}, \ndphi^{*}} \pr{z \gvn \ndphi}}{\pr{z \gvn \ndphi^{*}}}}{z},
\end{equation}
which is the very heart of \Chippr, also given as \Eq{eqn:final} and in the form of interpretive dance \citep{malz_probabilistic_2019}.
This in effect replaces the implicit prior with the sampled model hyperparameters, thereby converting the \pzip s into likelihoods in order to obtain unbiased posteriors.

\subsection{Model Limitations}
\label{sec:limitations}

Finally, we explicitly review the assumptions made by this approach, which are as follows:
\begin{enumerate}
	\item Photometric measurements of galaxies are statistically independent Poisson draws from the set of all galaxies such that \Eq{eqn:indiedat} and \Eq{eqn:indie} hold.
	\item We take the reported \pzip s to be accurate, free of model misspecification; 
	draws thereof must not be inconsistent with the distribution of photometry and redshifts.
	Furthermore, we must be given the implicit prior $\ndphi^{*}$ used to produce the \pzip s.
	\item We must assume a hyperprior distribution $\pr{\ndphi}$ constraining the underlying probability distribution of the hyperparameters, which is informed by our prior beliefs about the true redshift distribution function.
\end{enumerate}

These assumptions have known limitations.  
First, the photometric data are not a set of independent measurements; 
the data are correlated not only by the conditions of the experiment under which they were observed (instrument and observing conditions) but also by redshift covariances resulting from physical processes governing underlying galaxy spectra and their relation to the redshift distribution function.
Second, the reported \pzip s may not be trustworthy; 
there is not yet agreement on the best technique to obtain \pzpdf s, and the implicit prior may not be appropriate or even known to us as consumers of \pzip s.  
Third, the hyperprior may be quite arbitrary and poorly motivated if the underlying physics is complex, and it can only be appropriate if our prior beliefs about \nz\ are accurate.

Furthermore, in Section~\ref{sec:prob}, we have made an assumption of \textit{support}, meaning the model $\pr{z, \data \gvn \ndphi}$ has mutual coverage with the parameter values that real galaxies can take.  
In other words, any probability distribution over the $(z, \data)$ space must be nonzero where real galaxies can exist. 
Additionally, the hyperprior $\pr{\ndphi}$ must be nonzero at the hyperparameters $\ndphi^{\dagger}$ of the true redshift density function \nz.
This assumption cannot be violated under the experimental design of Section~\ref{sec:forward}, but it is not generically guaranteed when performing inference on real data;
thus the chosen $\pr{z, \data \gvn \ndphi^{*}}$ and $\pr{\ndphi}$ must be sufficiently general as to not rule out plausible areas of parameter space.

\section{Methods \& Data}
\label{sec:application}

Here we describe the method by which we demonstrate the \Chippr\ model.
In \Sect{sec:exp}, we outline the implementation of the \chippr\ code.
In \Sect{sec:mock}, we outline the procedure for emulating mock \pzip s.

\subsection{Implementation}
\label{sec:exp}

We implement the \Chippr\ model in code in order to perform tests of its validity and to compare its performance to that of traditional alternatives.
In \Sect{sec:mcmc}, we describe the publicly available \chippr\ library.
In \Sect{sec:sheldon}, we introduce the alternative approaches evaluated for comparison with \Chippr.
In \Sect{sec:diag}, we describe the diagnostic criteria by which we assess estimators of \nz.

\subsubsection{Code}
\label{sec:mcmc}

\chippr\ is a \python\ 2 library\footnote{\url{https://github.com/aimalz/chippr}} that includes an implementation of the \Chippr\ model as well as an extensive suite of tools for comparing \Chippr\ to other approaches.

Though there are plans for future expansion to more flexible parameterizations, the current version of \chippr\ uses a log-space piecewise constant parameterization
\begin{equation}
\label{eqn:logstepfunc}
f(z; \ndphi) = \exp[\phi^{k}]\ \mathrm{if}\ z^{k} < z < z^{k+1}
\end{equation}
for \nz\ and every \pzpdf, satisfying
\begin{equation}
\label{eqn:logstepfuncnorm}
\sum_{k=1}^{K} \exp[\phi^{k}] \delta z^{k} = 1
\end{equation}
with $K$ bins of width $\delta z^{1}, \dots, \delta z^{K}$ defined by endpoints $z^{0}, \dots, z^{K}$.
Thus each $\pr{z \gvn \data_{j}} = f(z; \ndphi_{j})$ has parameters $\ndphi_{j}$ that are defined in the same basis as those of \nz.
To infer the full log-posterior distribution $\ln[\pr{\ndphi \gvn \{\data_{j}\}}]$, one must provide a plaintext file with $K+1$ redshift bin endpoints $\{z_{k}\}$, the parameters $\ndphi^{*}$ of the implicit log-prior, and the parameters $\{\ndphi_{j}\}$ of the log-posteriors $\ln[\pr{z \gvn \data_{j}, \ndphi^{*})}$.

The \emcee \citep{foreman-mackey_emcee_2013} implementation of ensemble sampling is used to sample the full log-posterior of \Eq{eqn:final}. 
\chippr\ accepts a configuration file of user-specified parameters, among them the number $W$ of walkers.
At each iteration $i$ and for each walker, a proposal distribution $\hat{\ndphi}_{i}$ is drawn from the log-prior distribution and evaluated for acceptance to or rejection from the full log-posterior distribution.


The resulting output includes $\frac{I_{0}}{s}$ accepted samples $\ndphi_{i}$ for a pre-specified chain thinning factor $s$ and their full posterior probabilities $\pr{\ndphi_{i} \gvn \{\data_{j}\}}$, as well as the autocorrelation times and acceptance fractions calculated for each element of $\ndphi$, divided into separate files before and after the completion of the burn-in phase, as defined by the Gelman-Rubin statistic \citep{gelman_inference_1992}.  

\subsubsection{Alternative approaches for comparison}
\label{sec:sheldon}

In this study, we compare the results of \Eq{eqn:fullpost} to those of the two most common approaches to estimating \nz\ from a catalog of \pzip s: 
the distribution $n(z_{\mathrm{max}})$ of the redshifts at maximum posterior probability
\begin{equation}
\label{eqn:mmap}
f^{MMAP}(z; \hat{\ndphi}) = \sum_{j=1}^{J}\ \delta(z, \mathrm{mode}[\pr{z \gvn \data_{j}, \ndphi^{*}}])
\end{equation}
(i.e. the distribution of modes of the \pzip s) and the stacked estimator of \Eq{eqn:stacked}, which can be rewritten as 
\begin{equation}
\label{eqn:stacked}
f^{stack}(z; \hat{\ndphi}) = \sum_{j=1}^{J}\ \pr{z \gvn \data_{j}, \ndphi^{*}}
\end{equation}
in terms of the \pzip s we have.
These two approaches have been compared to one another by \citet{hildebrandt_cfhtlens:_2012}, \citet{benjamin_cfhtlens_2013}, and \citet{asorey_galaxy_2016} in the past but not to \Chippr.

Point estimation converts the implicit \pz\ posteriors $\pr{z \gvn \data_{j}, \ndphi^{*}}$ into delta functions with all probability at a single estimated redshift.  
Some variants of point estimation choose this single redshift to be that of maximum a posteriori probability $\mathrm{mode}[\pr{z \gvn \data_{j}, \ndphi^{*}}]$ or the expected value of redshift $\langle z \rangle = \integral{z \pr{z \gvn \data_{j}, \ndphi^{*}}}{z}$.
\citet{tanaka_photometric_2018} directs attention to deriving an optimal point estimate reduction of a \pzpdf, but since the purpose of this paper is to compare against the most established alternative estimators of \nz, its use will be postponed until a future study.
Stacking these modified \pzip s leads to the marginalized maximum a posteriori (MMAP) estimator and the marginalized expected value (MExp) estimator, though only the former is included in this study since the latter has fallen out of favor in recent years\footnote{And for good reason!  Consider a bimodal \pzpdf; its expected value may very well fall in a region of very low probability, yielding a less probable point estimate than the point at which either peak achieves its maximum.}.

It is worth discussing the relationship between point estimation and stacking.  
When the point estimator of redshift is equal to the true redshift, stacking delta function \pzpdf s will indeed lead to an accurate recovery of the true redshift distribution function.  
However, stacking is in general applied indiscriminately to broader \pzpdf s and imperfect point estimators of redshift.  
It is for these reasons that alternatives are considered here.

A final estimator of the hyperparameters is the maximum marginalized likelihood estimator (MMLE), the value of $\ndphi$ maximizing the log posterior given by \Eq{eqn:final} using any optimization code. 
The MMLE can be obtained in substantially less time than enough samples to characterize the full log-posterior distribution of \nz.
However, the MMLE yields only a point estimate of \nz\ rather than characterizing the full log-posterior on $\ndphi$, and it does not escape the dependence on the choice of hyperprior distribution.  
Furthermore, derivatives will not in general be available for the full posterior distribution, restricting optimization methods used, and, as is true for any optimization code, there is a risk of numerical instability.

\subsubsection{Performance metrics}
\label{sec:diag}

The results of the computation described in \Sect{sec:exp} are evaluated for accuracy on the basis of some quantitative measures.  
Beyond visual inspection of samples, we calculate summary statistics to quantitatively compare different estimators' precision and accuracy.  
Since MCMC samples of these hyperparameters are Gaussian distributions, we can quantify the breadth of the distribution for each hyperparameter using the standard deviation regardless of whether the true values are known.  

In simulated cases where the true parameter values are known, we calculate the Kullback-Leibler divergence (KLD), given by 
\begin{equation}
\label{eqn:kl}
KL_{\ndphi,\ndphi^{\ddagger}} = \integral{\pr{z \gvn \ndphi} \ln \left[ \frac{\pr{z \gvn \ndphi}}{\pr{z \gvn \ndphi^{\dagger}}} \right]}{z} ,
\end{equation}
which measures a distance from parameter values $\ndphi$ to true parameter values $\ndphi^{\dagger}$.  
The KLD is a measure of the information loss, in units of nats, due to using $\ndphi$ to approximate the true $\ndphi^{\dagger}$ when it is known.
A detailed exploration of the KLD may be found in the Appendix to \citet{malz_approximating_2018}.

\subsection{Validation on mock data}
\label{sec:mock}

\begin{figure*}
	\begin{center}
		\includegraphics[width=0.7\textwidth]{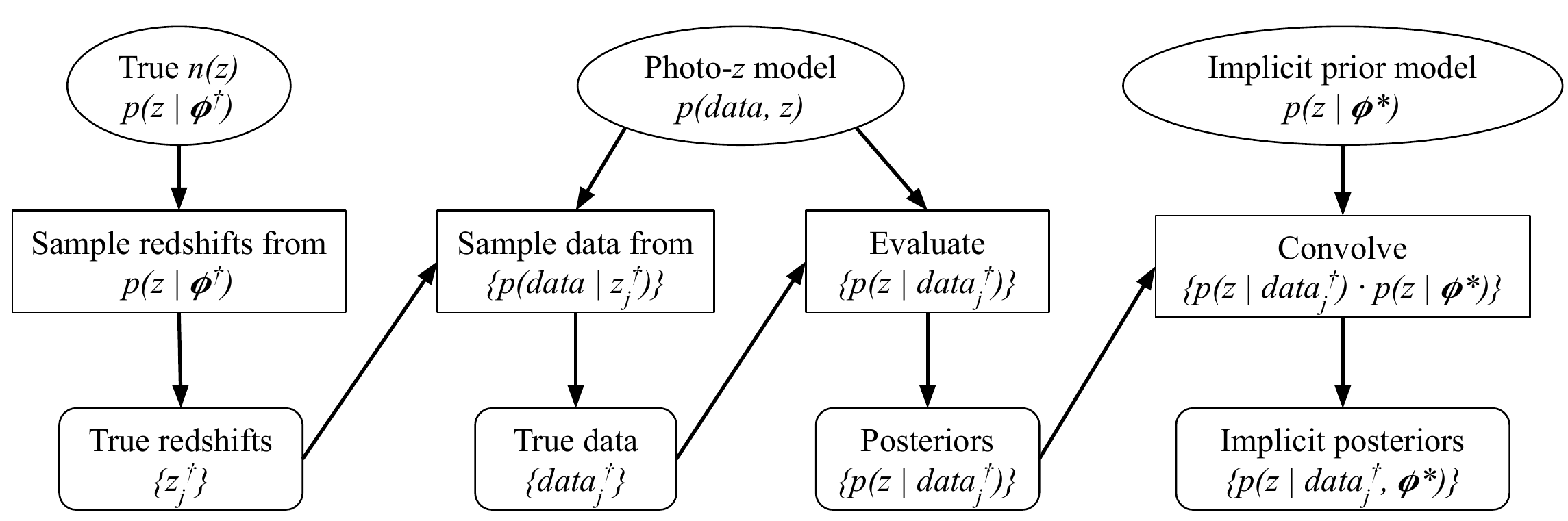}
		\caption{
			A flow chart illustrating the forward model used to generate mock data in the validation of \Chippr, as described in \Sect{sec:forward}.
			Ovals indicate a quantity that must be chosen in order to generate the data, rectangles indicate an operation we perform, and rounded rectangles indicate a quantity created by the forward model.
			Arrows indicate the inputs and outputs of each operation performed to simulate mock \pzip\ catalogs.
		}
		\label{fig:flowchart}
	\end{center}
\end{figure*}

We compare the results of \Chippr\ to those of stacking and the histogram of \pzip\ maxima (modes) on mock data in the form of catalogs of emulated \pzip s generated via the forward model discussed in \Sect{sec:forward}.
\Fig{fig:flowchart} illustrates the implementation of the forward model, defined by the much simpler \Fig{fig:pgm}, used for validating the method presented here.
The irony of a simple model and complex validation procedure is not lost on the authors.

\Fig{fig:flowchart} outlines the four phases of the generative model, which uses a total of three inputs.
The experimental design requires our choice of true values $\phi^{\dagger}$ of the hyperparameters governing \nz, a \pz\ model $\pr{z, \data}$ defining the space of redshift and photometry, and prior values $\phi^{*}$ of the hyperparameters of \nz.
In the first phase, we sample $J = 10^{4}$ redshifts $z_{j}^{\dagger} \sim \pr{z \gvn \phi^{\dagger}}$.
In the second phase, we evaluate the \pz\ model at those redshifts, yielding a set of $J$ likelihoods $\pr{\data \gvn z_{j}^{\dagger}}$, from which we then sample data $\data_{j}^{\dagger} \sim \pr{\data \gvn z_{j}^{\dagger}}$ for each galaxy.
In the third phase, we evaluate the \pz\ model at that data to obtain $J$ posteriors $\pr{z \gvn \data_{j}^{\dagger}}$.
In the fourth phase, we convolve the posteriors with the chosen prior $\pr{z \gvn \phi^{*}}$, yielding implicit posteriors $\pr{z \gvn \data_{j}^{\dagger}, \phi^{*}}$.

The true redshift distribution used in these tests is a particular instance of the gamma function
\begin{equation}
\label{eqn:gamma}
n^{\dagger}(z) = \frac{1}{2 c_{z}} \left(\frac{z}{c_{z}}\right)^{2}\ \exp\left[-\frac{z}{c_{z}}\right]
\end{equation}
with $c_{z} = 0.3$, because it has been used in forecasting studies for \des\ and \lsst.

\begin{figure}
	\begin{center}
		\includegraphics[width=0.45\textwidth]{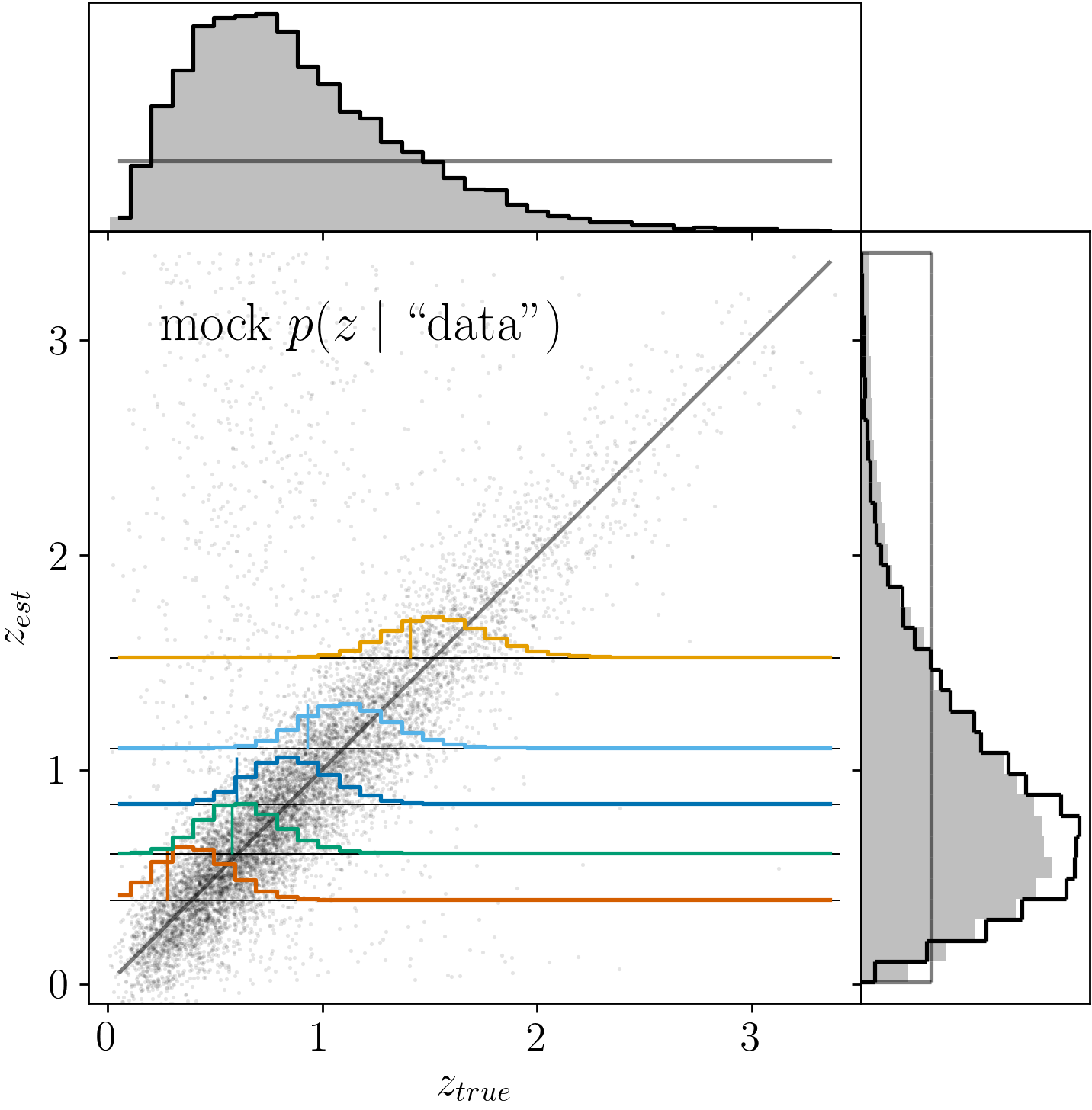}
		\caption{
			The joint probability space of true and estimated redshift for the three concerning \pz\ systematics at the level of the \lsst\ requirements: 
			intrinsic scatter, uniformly distributed catastrophic outliers, and bias.
			The main panel shows samples (black points) in the space of mock data and redshift, akin to the standard scatterplots of true and estimated redshift, the $z_{\mathrm{spec}} = z_{\mathrm{phot}}$ diagonal (gray line), and posterior probabilities evaluated at the given estimated redshift (colored step functions).
			The insets show marginal histograms (light gray) in each dimension, that can be compared with the true \nz\ used to make the figure (black) to see the effect of these systematics, as well as the implicit prior (dark gray).
		}
		\label{fig:mega_scatter}
	\end{center}
\end{figure}

The mock data emulates the three sources of error of highest concern to the \pz\ community that are explored in detail later in this section: intrinsic scatter (\Sect{sec:scatter}), catastrophic outliers (\Sect{sec:outliers}), and canonical bias (\Sect{sec:bias}).
\Fig{fig:mega_scatter} illustrates these three effects simultaneously at the tolerance of \lsst\ for demonstrative purposes, harking back to Figure~\ref{fig:pedagogical_scatter}.

The hyperprior distribution chosen for these tests is a multivariate normal distribution with mean $\vec{\mu}$ equal to the implicit prior $\ndphi^{*}$ and covariance
\begin{equation}
\label{eqn:priorcov}
\Sigma_{k,k'} = q\ \exp[-\frac{e}{2}\ (\bar{z}_{k}-\bar{z}_{k'})^{2}]\ +\ t\delta(k,k')
\end{equation}
inspired by one used in Gaussian processes, where $k$ and $k'$ are indices ranging from $1$ to $K$ and $q=1.0$, $e=100.0$, and $t=q\cdot10^{-5}$ are constants chosen to permit draws from this prior distribution to produce shapes similar to that of a true $\tilde{\ndphi}$.  
We adapt the full log-posterior of \Eq{eqn:final} to the chosen binning of redshift space.

The sampler is initialized with $W=100$ walkers each with a value chosen from a Gaussian distribution of identity covariance around a sample from the hyperprior distribution.  

\section{Results}
\label{sec:alldata}

Here, we compare the results of the \Chippr\ methodology with those of established \nz\ estimators under the three traditional measures of \pz\ uncertainty one at a time:
\Sect{sec:scatter} concerns the redshift-dependent intrinsic scatter, \Sect{sec:outliers} concerns realistically complex catastrophic outlier populations, and \Sect{sec:bias} concerns the canonical bias in the mean redshift.

\subsection{Intrinsic scatter}
\label{sec:scatter}

\begin{figure*}
	\begin{center}
	\includegraphics[width=0.45\textwidth]{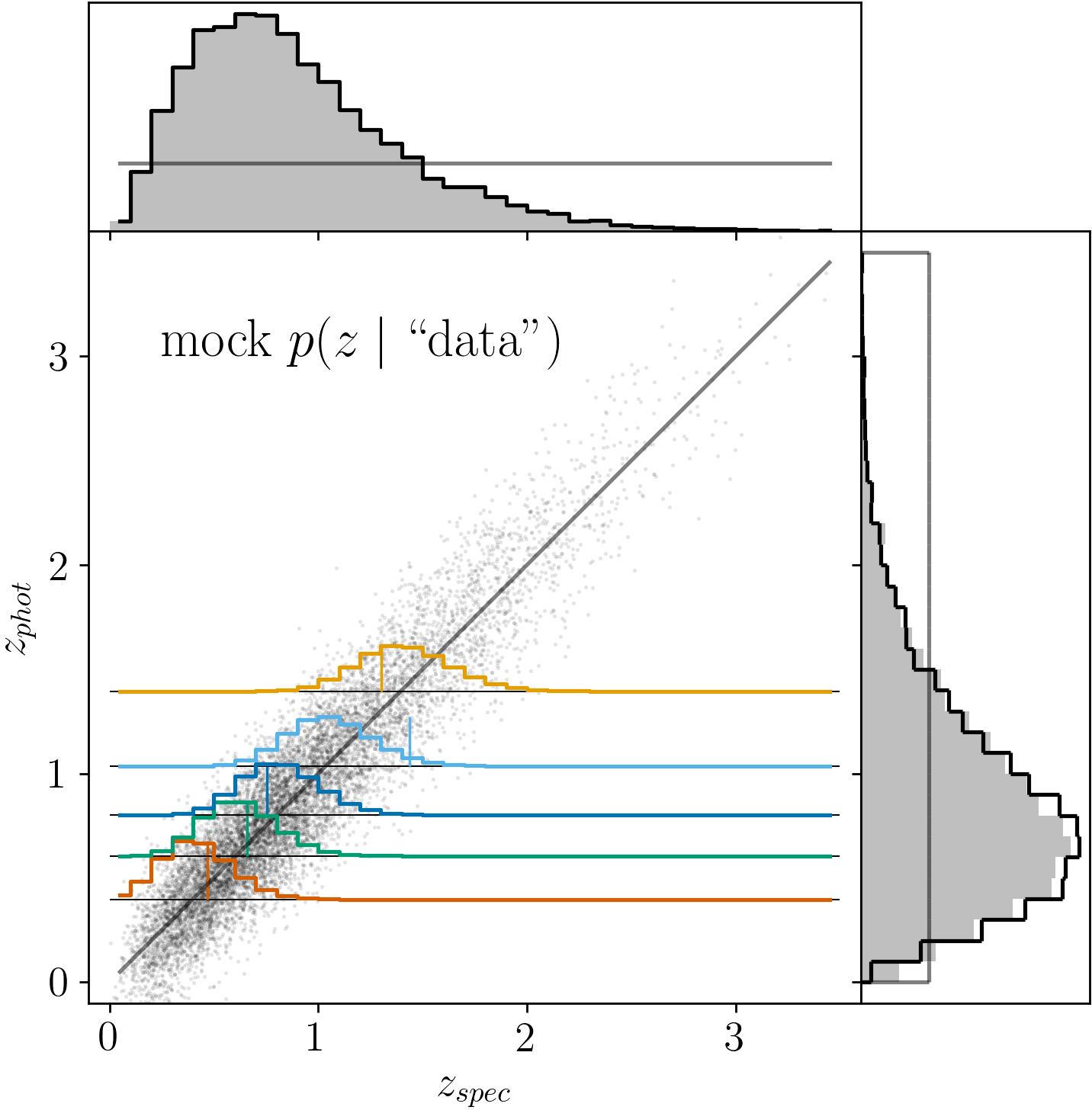}
	\includegraphics[width=0.45\textwidth]{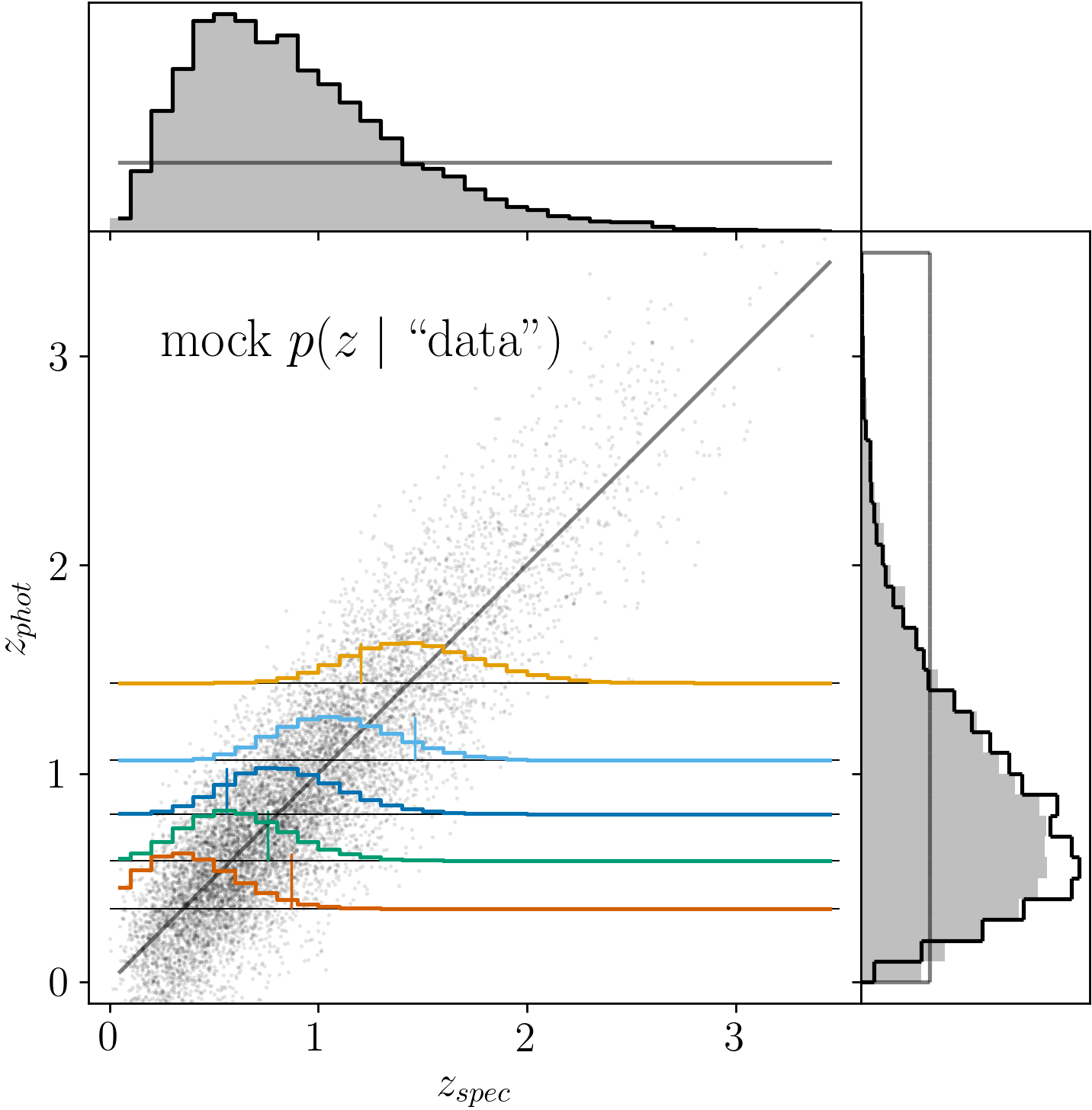}
	\caption{
		Examples of mock \pzpdf s generated with intrinsic scatter at the \lsst\ requirements (left) and twice the \lsst\ requirements (right), including samples from the probability space of true and observed redshift (black points), \pzpdf s (colored step functions), and the true redshifts of the example \pzpdf s (colored vertical lines).
		A histogram (light gray) of points in each dimension is shown in the respective inset, with the true redshift distribution (black) and implicit prior (dark gray).
	}
	\label{fig:pzs-scatter}
	\end{center}
\end{figure*}

\Fig{fig:pzs-scatter} shows some examples of \pzpdf s generated with only the systematic of intrinsic scatter, at the level of the \lsst\ requirements on the left and twice that on the right.
One can see that the histogram of redshift estimates is broader than that of true redshifts, and that the effect is substantially more pronounced by just doubling the intrinsic scatter from the level of the \lsst\ requirements.

\begin{figure*}
	\begin{center}
		\includegraphics[width=0.45\textwidth]{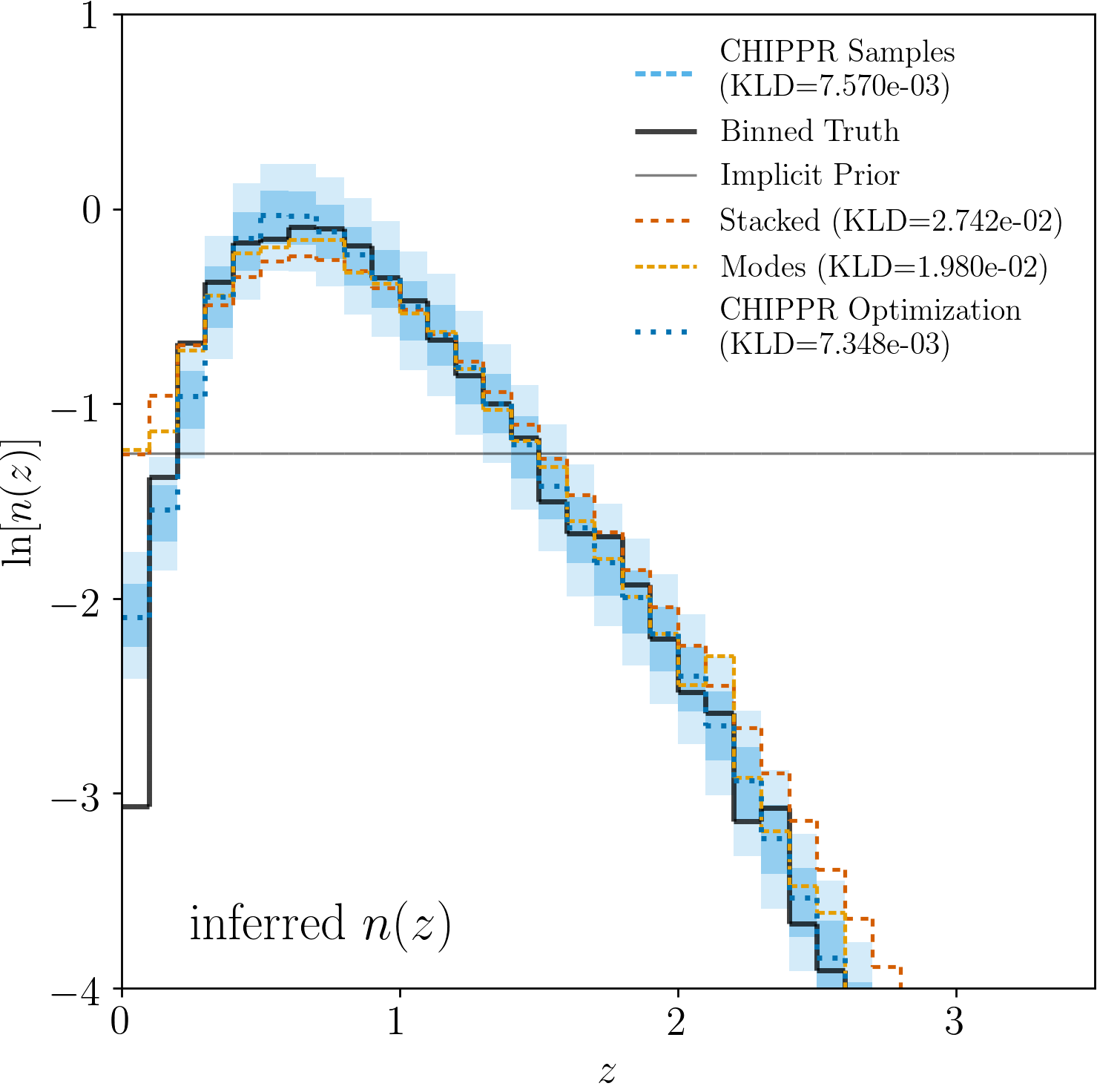}
		\includegraphics[width=0.45\textwidth]{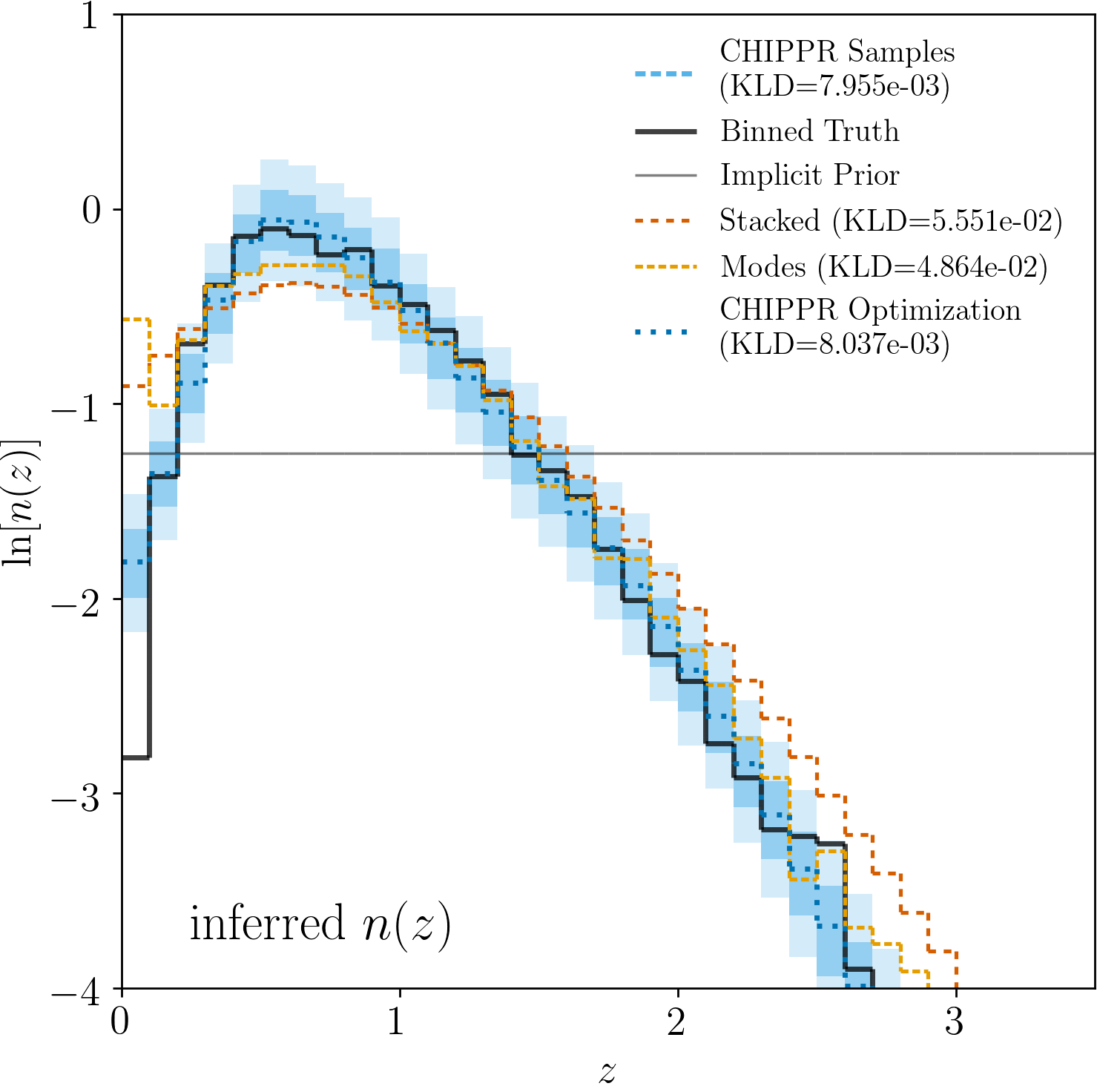}
	\caption{
		The results of \Chippr\ (samples in light blue and optimization in dark blue) and the alternative approaches (the stacked estimator in red and the histogram of modes in yellow) on \pzpdf s with intrinsic scatter of the \lsst\ requirements (left) and twice that (right), with the true redshift density (black curve) and implicit prior (gray curve).
		\Chippr\ is robust to intrinsic scatter, but the alternatives suffer from overly broad \nz\ estimates that worsen with increasing intrinsic scatter.
	}
	\label{fig:results-scatter}
	\end{center}
\end{figure*}

\Fig{fig:results-scatter} shows the \nz\ recovered by \Chippr\ and the alternative approaches.
As expected, the estimates of \nz\ based on the modes of the \pzpdf s and stacking are broader than the marginalized maximum likelihood estimator from \chippr, with more broadening as the intrinsic scatter increases.
\Chippr's \mmle\ is robust to intrinsic scatter and is unaffected by increased intrinsic scatter, though the \Chippr\ posterior distribution on the redshift distribution is itself broader for the higher intrinsic scatter case than for the \lsst\ requirements.
The broadening of the alternative estimators corresponds to a loss of 3-4 times as many nats of information about \nz\ for the \lsst\ requirements relative to the \mmle\ of \Chippr.

\subsection{Catastrophic outliers}
\label{sec:outliers}

As was covered in \Sect{sec:intro}, catastrophic outliers tend to be distributed non-uniformly across the space of observed and true redshift.
However, the \lsst\ requirements do not specify details for a distribution of outliers to which they were tuned, and it is still instructive to examine the impact of uniform outliers on the inference of \nz, so we begin by addressing uniformly distributed outliers before considering more realistic outlier distributions.

\begin{figure}
	\begin{center}
	\includegraphics[width=0.45\textwidth]{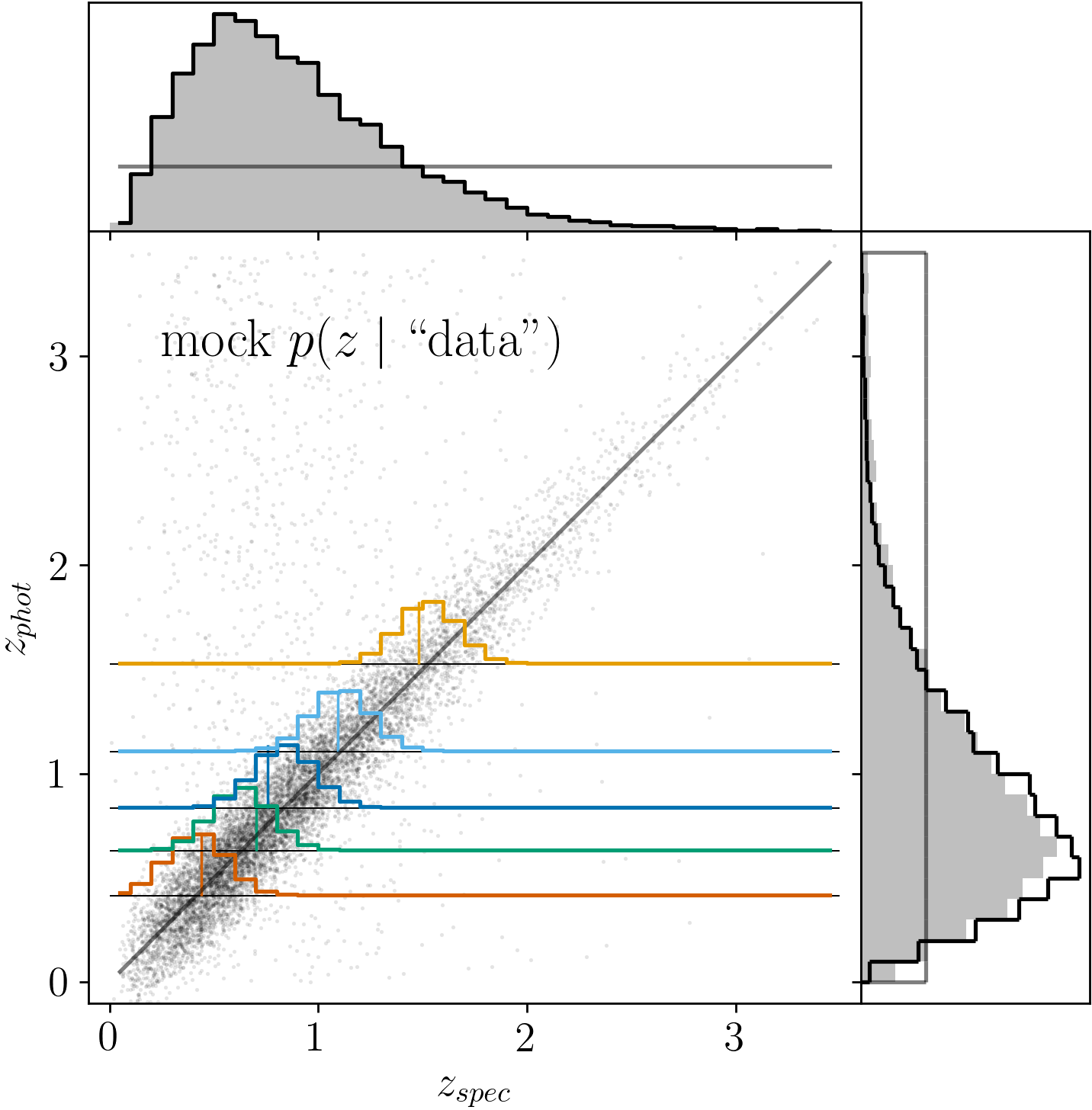}\\
	\includegraphics[width=0.45\textwidth]{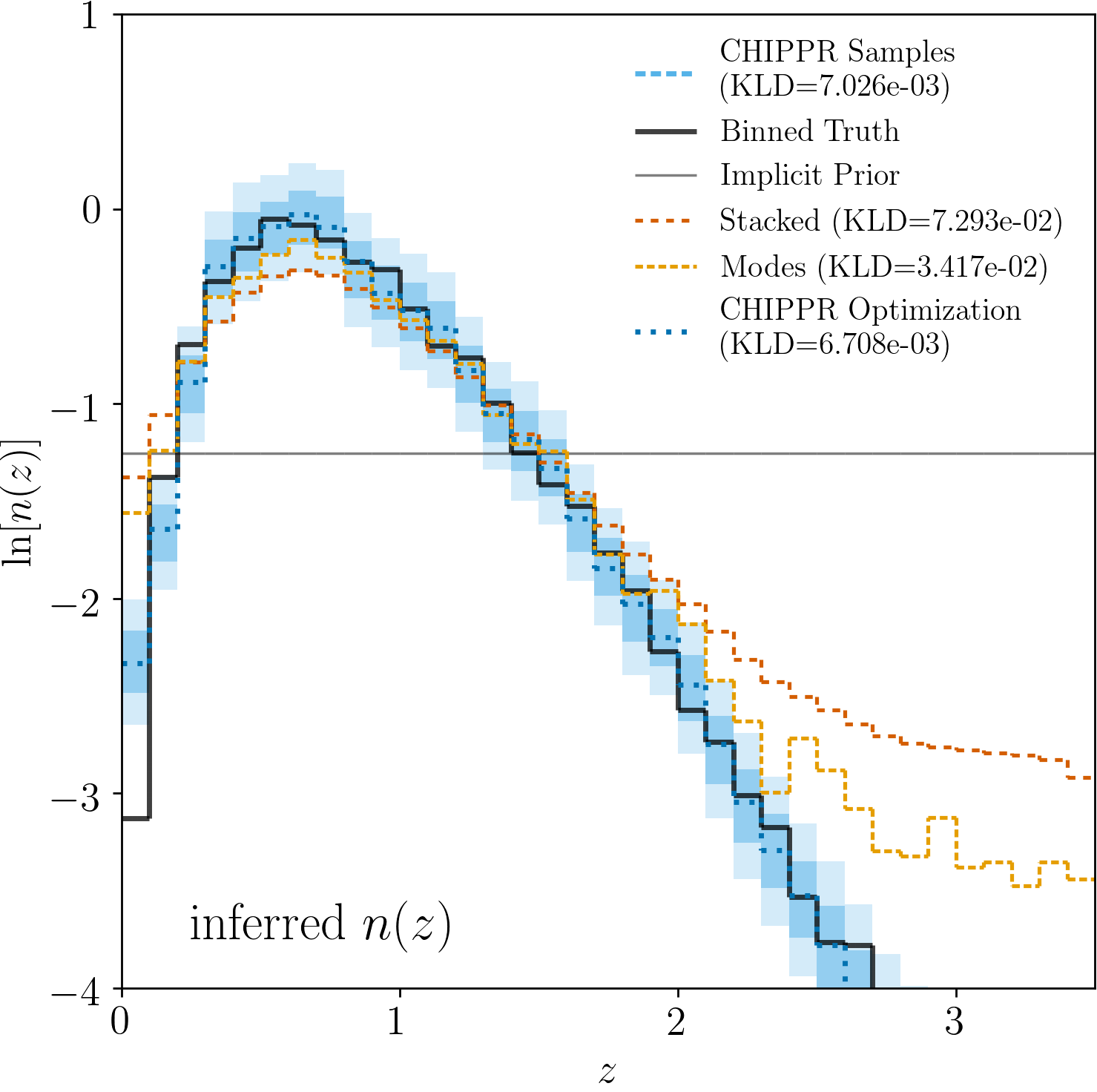}
	\caption{
		Top: Examples of \pzpdf s with a uniformly distributed catastrophic outlier population at the level of the \lsst\ requirements, including samples from the probability space of true and observed redshift (black points), \pzpdf s (colored step functions), and the true redshifts of the example \pzpdf s (colored vertical lines), with marginal histograms (light gray) for each dimension with the true redshift distribution (black) and implicit prior (dark gray) in the insets.
		Bottom: The results of \Chippr\ (samples in light blue, optimization in dark blue) and the alternative approaches (the stacked estimator in red, the histogram of modes in yellow) on \pzpdf s with uniformly distributed catastrophic outliers, with the true redshift density (black curve) and implicit prior (gray curve).
		The presence of the catastrophic outlier population broadens the histogram of modes and stacked estimator of the redshift distribution, but the result of \Chippr\ is unbiased.
	}
	\label{fig:uniform-outliers}
	\end{center}
\end{figure}

A uniformly distributed population of outliers was simulated by giving every sample in true redshift a $10\%$ chance of having an observed redshift drawn from a uniform distribution rather than the Gaussian about the true redshift.
Though this results in slightly less than the $10\%$ catastrophic outlier rate, it can be done independently of the definition of the standard deviation so was implemented for demonstrative purposes.
\Fig{fig:uniform-outliers} shows examples of \pzpdf s from a uniformly distributed outlier population at the level of the \lsst\ requirements (left) as well as the results of \Chippr\ and other \nz\ estimation methods (right).
The intrinsic scatter of the tests in this section does not increase with redshift as indicated in Table~\ref{tab:lsstsrd}
in order to isolate the effect of outliers, and is instead held at a constant $\sigma_{z} = 0.02$.

\Fig{fig:uniform-outliers} shows that at the level of the \lsst\ requirements, the alternative estimators are overly broad, whereas \Chippr's \mmle\ yields an unbiased estimate of \nz.
Further, the result of stacking is even broader than that of the histogram of modes, corresponding to ten times the information loss of \Chippr's \mmle, making it worse than the most naive reduction of \pzpdf s to point estimates.

When one thinks of the \pzpdf s of catastrophic outliers, however, what comes to mind is multimodal \pzpdf s, wherein reducing \pzpdf s to point estimates to make a standard scatterplot of the true and observed redshifts leads to substantial probability density off the diagonal.
These coordinated catastrophic outliers may be emulated in the joint probability space of true and estimated redshifts by using a mixture of the unbiased diagonal defined by the intrinsic scatter and an additional Gaussian in one dimension, with constant observed redshift for a template-fitting code and constant true redshift for a machine learning code.

\begin{figure*}
	\begin{center}
	\includegraphics[width=0.45\textwidth]{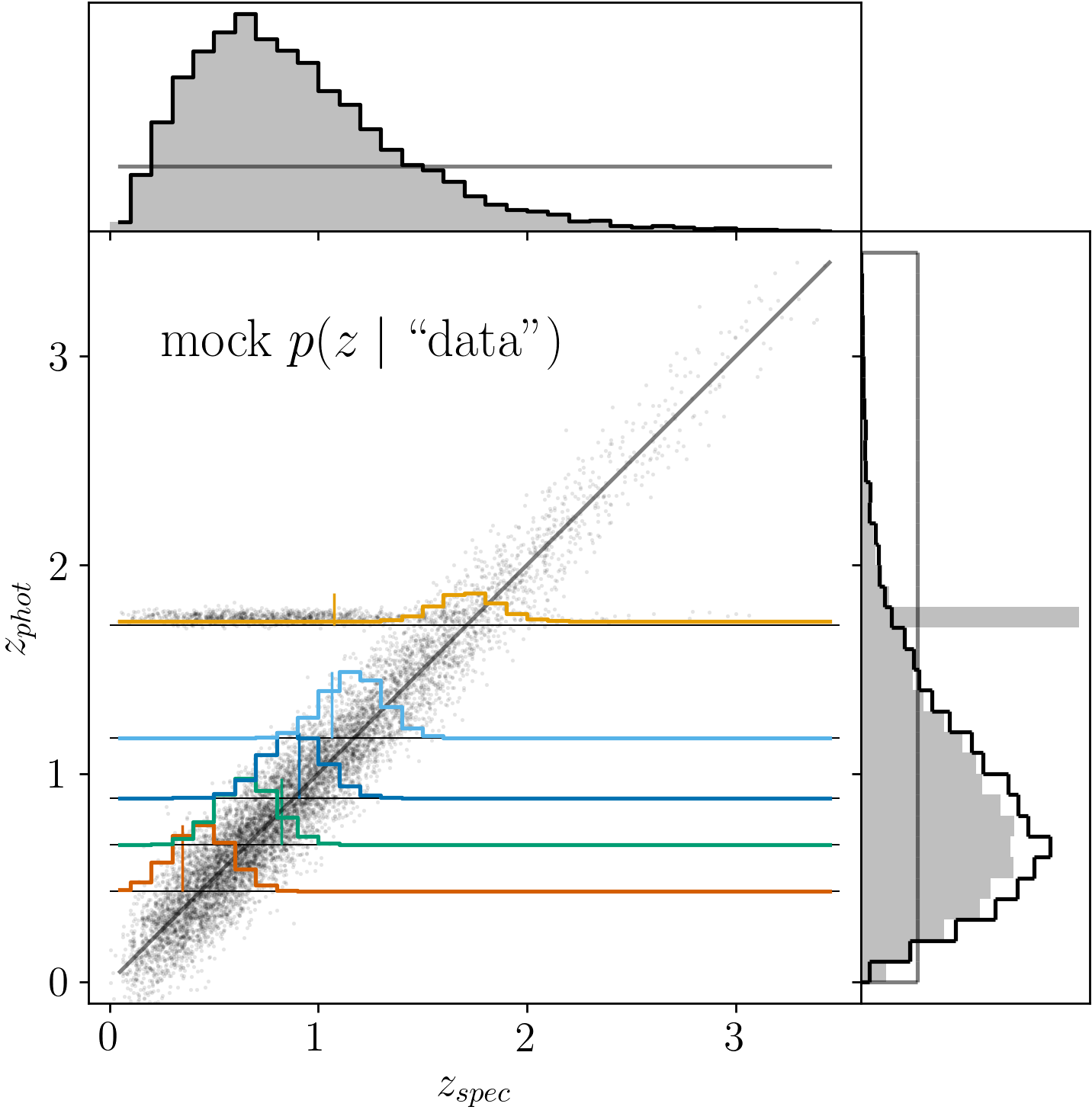}
	\includegraphics[width=0.45\textwidth]{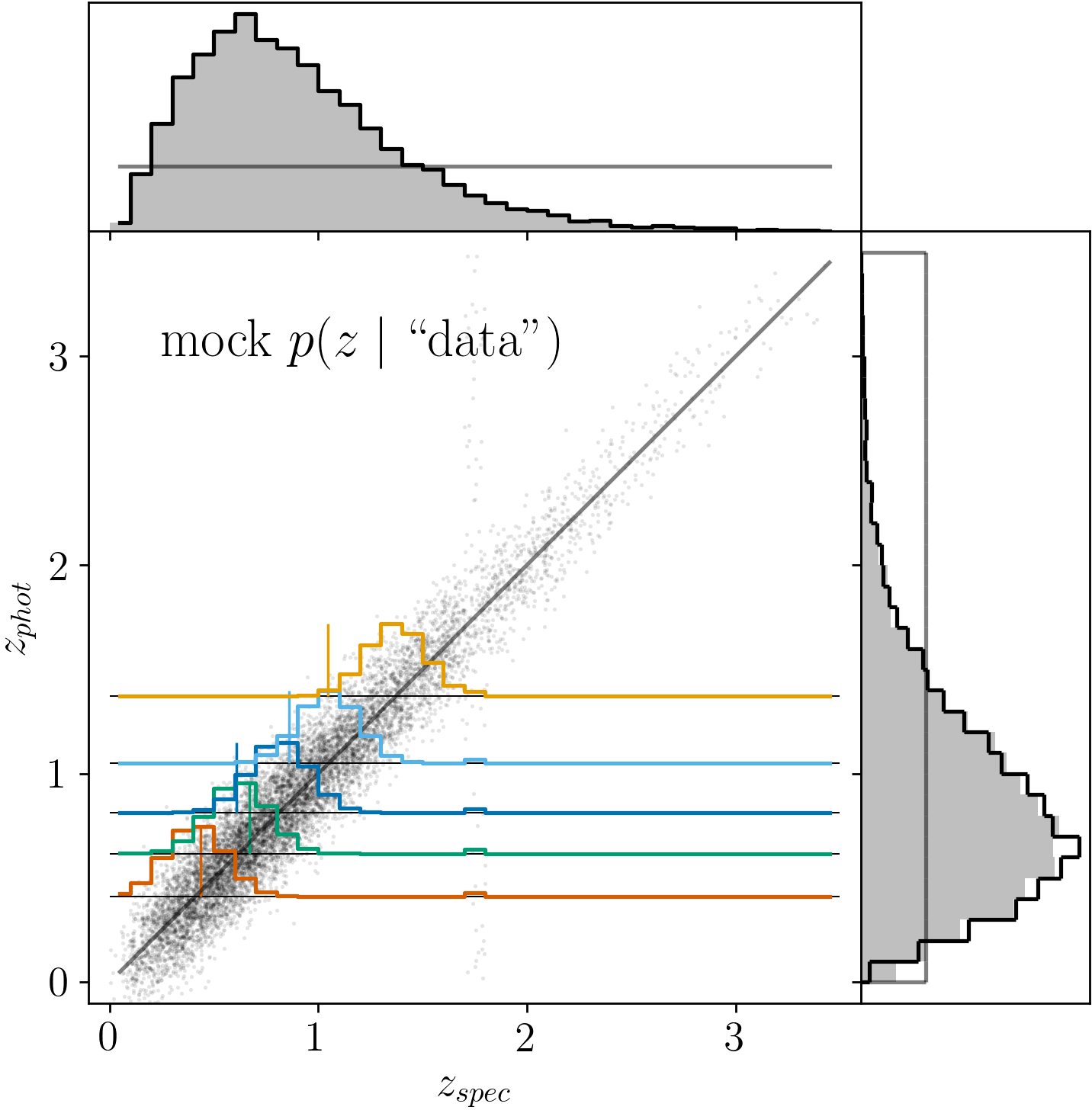}
	\caption{
		Examples of \pzpdf s with a catastrophic outlier population like that seen in template-fitting \pzpdf\ codes (left) and machine learning \pzpdf\ codes (right), including samples from the probability space of true and observed redshift (black points), \pzpdf s (colored step functions), and the true redshifts of the example \pzpdf s (colored vertical lines), with marginal histograms (light gray) for each dimension with the true redshift distribution (black) and implicit prior (dark gray) in the insets.		
	}
	\label{fig:nonuniform-outliers-data}
	\end{center}
\end{figure*}

In the case of a catastrophic outlier population like that anticipated of template-fitting codes, $10\%$ of all galaxies have their observed redshift at a particular value unrelated to their true redshift, illustrated in the left panel of \Fig{fig:nonuniform-outliers-data}.
This case is subject to the same caveat as the uniformly distributed outliers when it comes to the \lsst\ requirement.
It is less straightforward to emulate catastrophic outliers like those anticipated of a machine learning code, those that are truly multimodal.
The testing conditions here, illustrated in the right panel of \Fig{fig:nonuniform-outliers-data}, gives $10\%$ of galaxies at the redshift affected by outliers an observed redshift that is uniformly distributed relative to the true redshift, meaning that far fewer than $10\%$ of all galaxies in the sample are catastrophic outliers.

\begin{figure*}
	\begin{center}
	\includegraphics[width=0.45\textwidth]{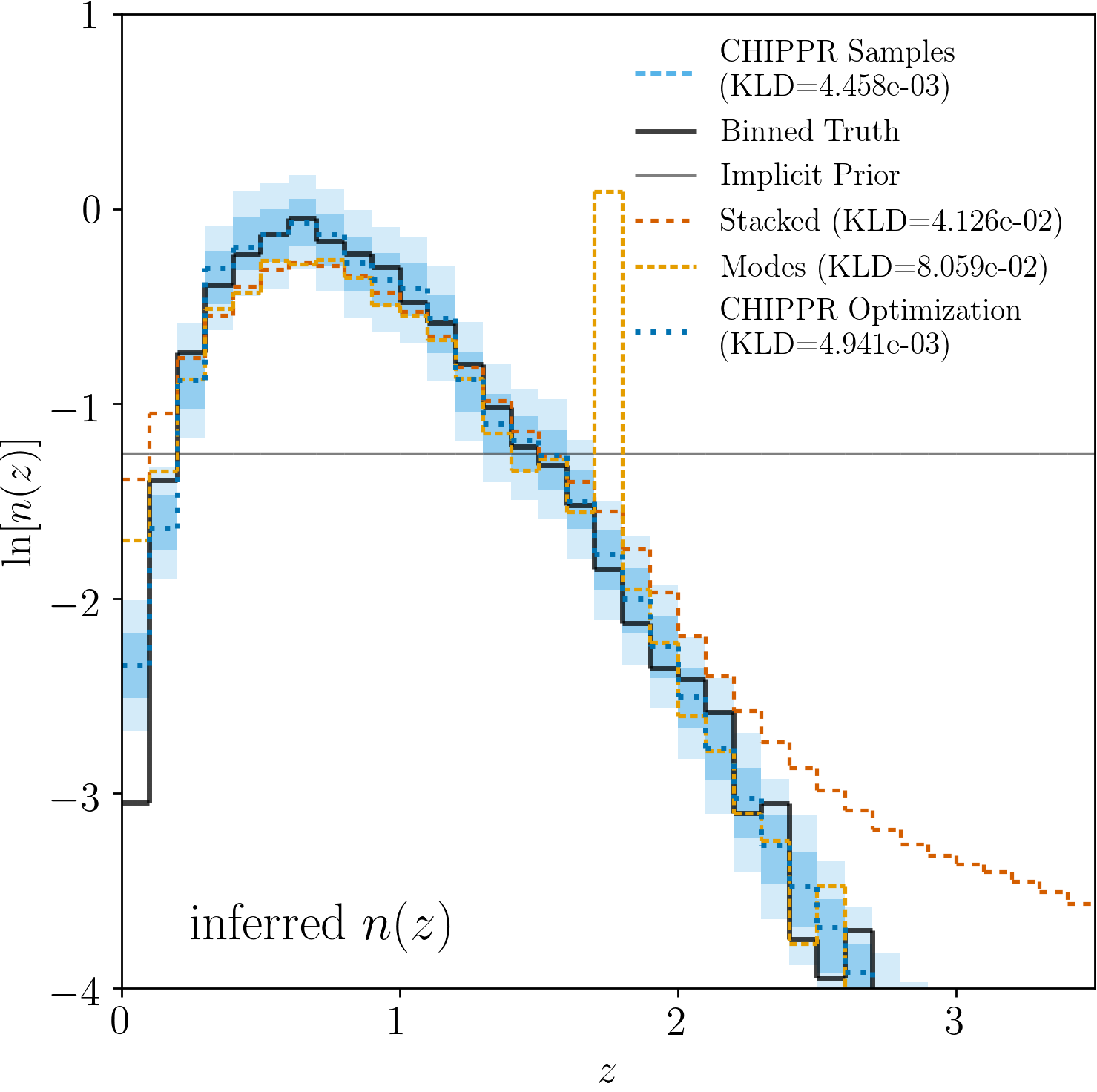}
	\includegraphics[width=0.45\textwidth]{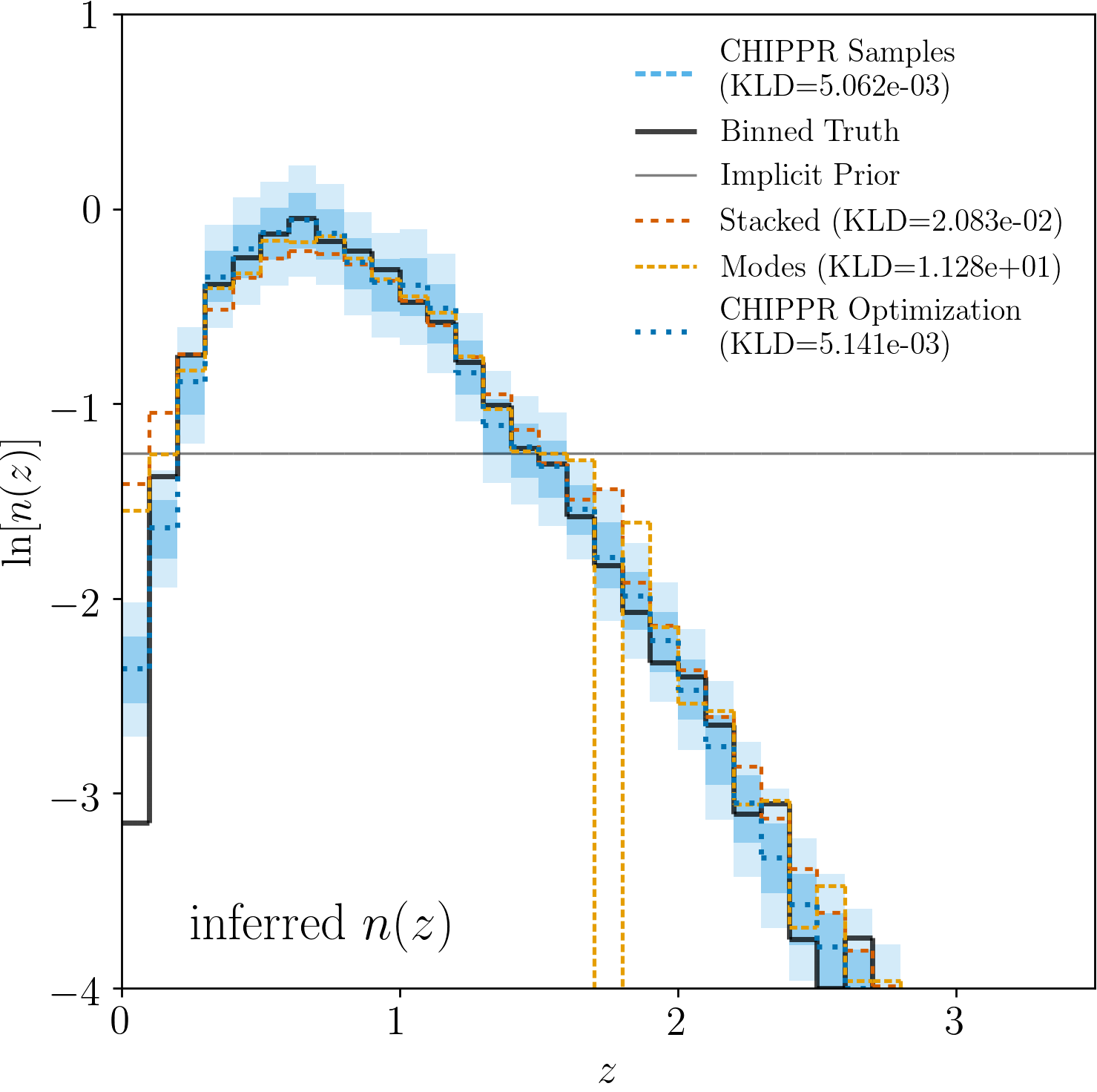}
	\caption{
		The results of \Chippr\ (samples in light blue and optimization in dark blue) and the alternative approaches (the stacked estimator in red, the histogram of modes in yellow) on \pzpdf s with catastrophic outliers like those seen in template-fitting \pzpdf\ codes (left) and machine learning \pzpdf\ codes (right) to the \lsst\ requirements, with the true redshift density (black curve) and implicit prior (gray curve).
		Though the histogram of modes is most sensitive to a catastrophic outlier population, the stacked estimator also overestimates \nz\ under (machine learning-like outliers) and beyond (template fitting-like outliers).
	}
	\label{fig:nonuniform-outliers-results}
	\end{center}
\end{figure*}

The results of \Chippr\ and the alternative estimators of \nz\ are presented in \Fig{fig:nonuniform-outliers-results}.
The most striking feature is that the histogram of modes is highly sensitive to both outlier populations, producing a severe overestimate in the case of an outlier population like those seen in template-fitting codes and a severe underestimate in the case of an outlier population like those seen in machine learning codes, corresponding to a twenty-fold loss of information compared to the \Chippr\ \mmle\ in both cases.
The effect on the stacked estimator of \nz\ is more subtle though still concerning.
In the case of outliers like those resulting from template-fitting, the stacked estimator is overly broad even without realistic intrinsic scatter, resulting in ten times the information loss compared to the \Chippr\ \mmle, and in the case of outliers like those resulting from machine learning, the stacked estimator features an overestimate at the redshift affected by the outlier population, resulting in about five times the information loss as the \Chippr\ \mmle.
The \Chippr\ \mmle, however, appears unbiased and withstands these effects, and the breadth of the distribution of samples of \nz\ is invariant.

\subsection{Canonical bias}
\label{sec:bias}

Systematic bias in \pz\ point estimates, is a concern for \lsst's cosmology results, for the same reasons explored in \citet{hoyle_dark_2018}.
This form of bias is typically summarized by a shift parameter $\Delta_{z} = (\langle \pr{z \gvn \hat{\ndphi}} \rangle - \langle \pr{z \gvn \ndphi^{\dagger}} \rangle)$ representing a difference between the first moment of the estimated redshift density function and that of the true redshift density function.
To distinguish other aforementioned manifestations of bias from this common form of bias, we refer to $\Delta_{z}$ as the \textit{canonical bias}.

\begin{figure}
	\begin{center}
	\includegraphics[width=0.45\textwidth]{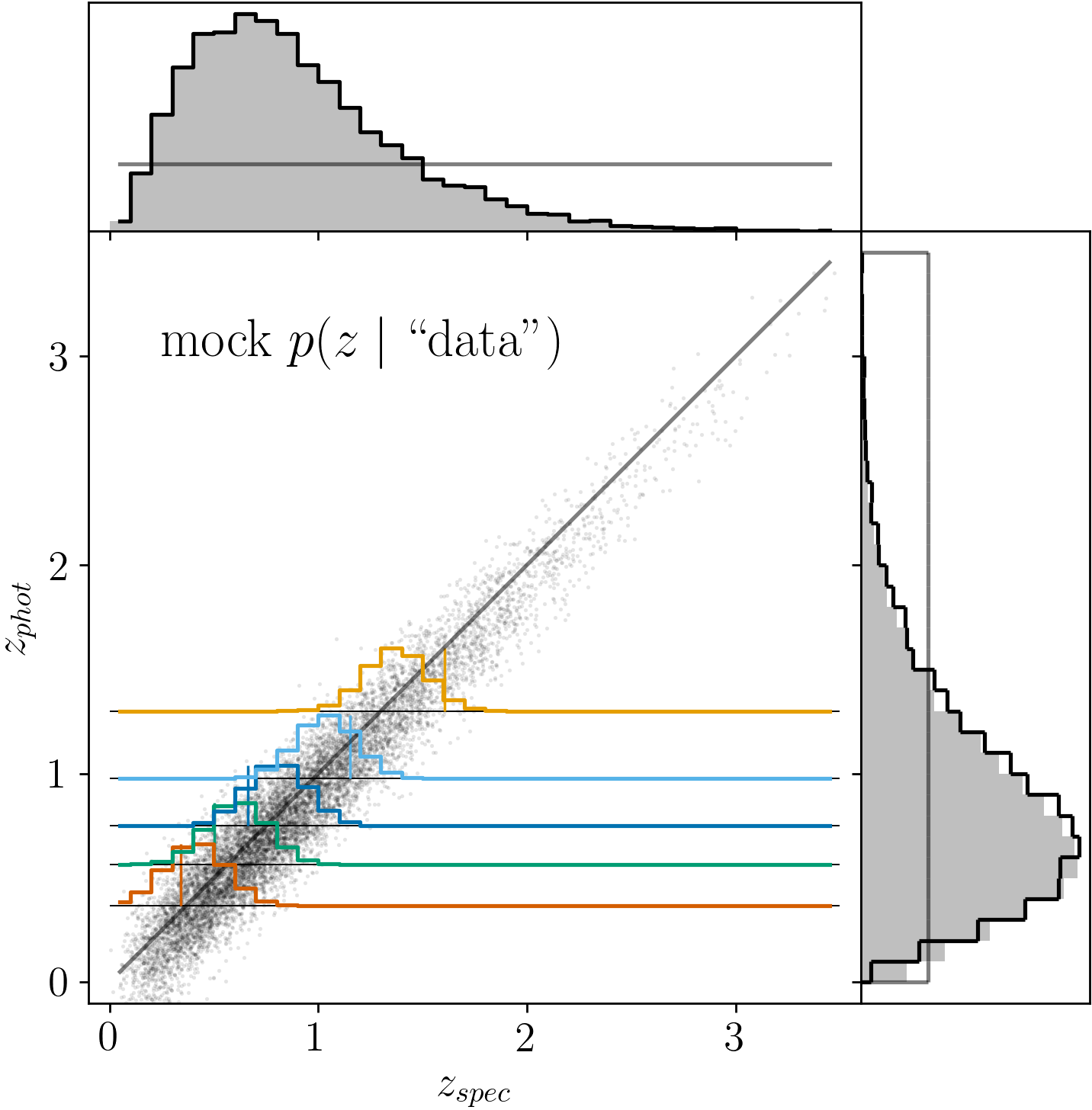}\\
	\includegraphics[width=0.45\textwidth]{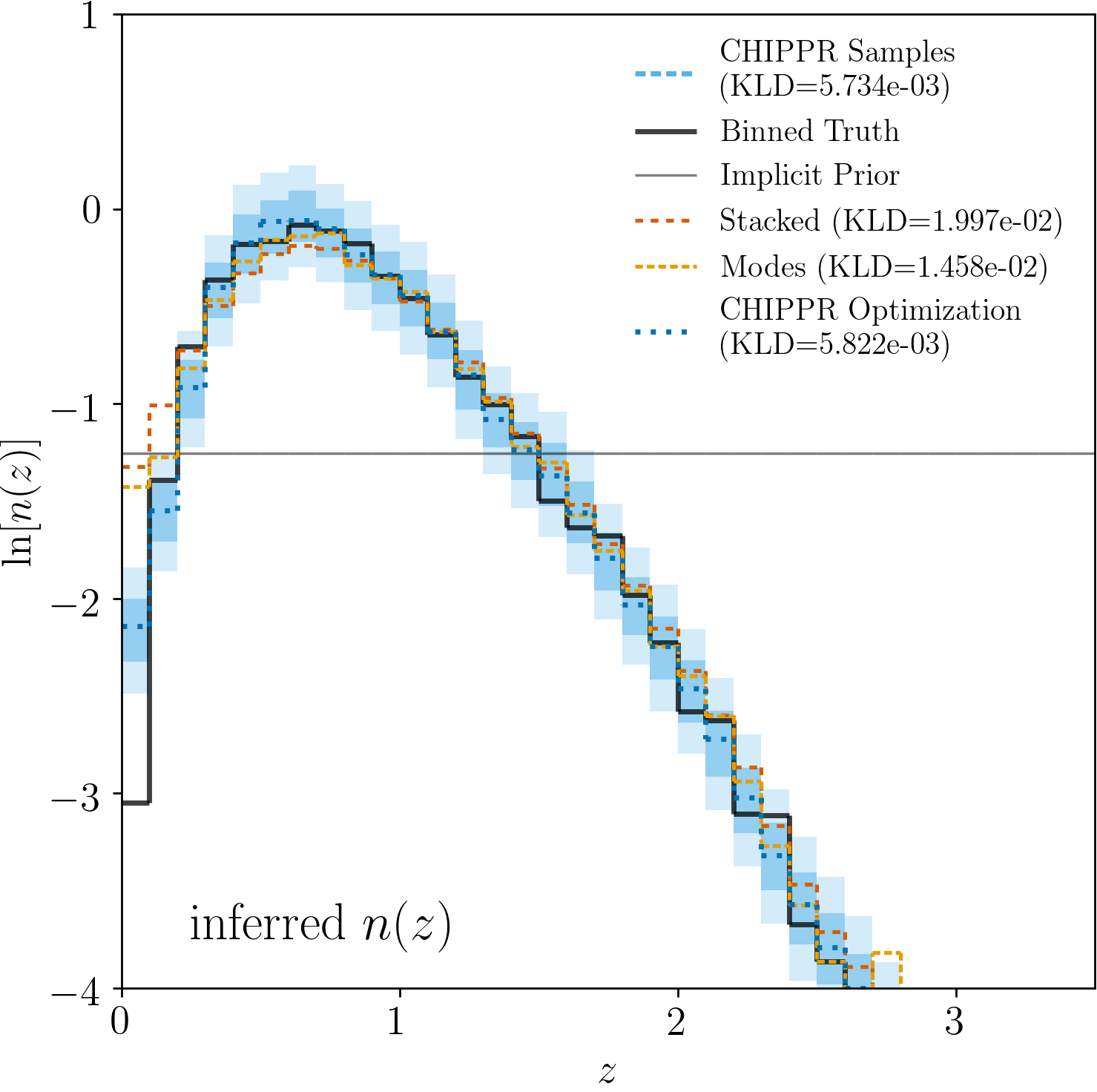}
	\caption{
		Top: Examples of \pzpdf s with ten times the bias of the \lsst\ requirements, including samples from the probability space of true and observed redshift (black points), \pzpdf s (colored step functions), and the true redshifts of the example \pzpdf s (colored vertical lines), with marginal histograms (light gray) for each dimension with the true redshift distribution (black) and implicit prior (dark gray) in the insets.
		Bottom: The results of \Chippr\ (samples in light blue, optimization in dark blue) and the alternative approaches (the stacked estimator in red, the histogram of modes in yellow) on \pzpdf s with ten times the bias of the \lsst\ requirements, with the true redshift density (black curve) and implicit prior (gray curve).
		The impact of bias at even ten times the level of the \lsst\ requirements is almost imperceptible on all estimators, though the \Chippr\ \mmle\ minimizes the information loss regardless.
	}
	\label{fig:bias}
	\end{center}
\end{figure}

In the context of \pzpdf s, the canonical bias represents an instance of model misspecification.
Consider that if the canonical bias were included in the framework of Figure~\ref{fig:pedagogical_scatter}, it could be trivially modeled out as a simple linear transformation of $z_{\mathrm{phot}} \to z_{\mathrm{phot}} - \Delta_{z} (1 + z_{\mathrm{phot}})$ of the $(z_{\mathrm{spec}}, z_{\mathrm{phot}})$ space.
Regardless, for completeness, a test at ten times the canonical bias of the \lsst\ requirements, with no redshift-dependent intrinsic scatter nor catastrophic outliers, is provided in \Fig{fig:bias}.

As expected based on self-consistency of the forward-modeled \pzpdf s, \Chippr\ is immune to linear bias of the form of $\Delta_{z}$.
Furthermore, the alternative estimators are only weakly affected, with information loss two and four times greater than that of the \Chippr\ \mmle\ for the histogram of modes and stacked estimator respectively.
(This general robustness may suggest that the canonical bias may not be the most relevant measure of performance of estimators of \nz.)

\section{Discussion}
\label{sec:results}

The experiments of \Sect{sec:alldata} quantify the influence on each estimator of \nz\ due to each of the canonical types of \pz\ error one at a time in isolation.
Now, we stress-test \Chippr\ by exploring the impact of the implicit prior, which has thus far not received much attention in the literature.
\Sect{sec:interim} demonstrates the sensitivity of \nz\ estimation methods to realistically complex implicit priors, and \Sect{sec:violations} demonstrates the consequences of mischaracterization of the implicit prior used to generate the \pzip\ catalog.
These results provide compelling motivation for the \pz\ community to prioritize the study of implicit priors of existing and developing \pzpdf\ techniques.

\subsection{Realistically complex implicit prior}
\label{sec:interim}

\chippr\ can handle any implicit prior with support over the redshift range where \nz\ is defined, but some archetypes of implicit prior are more likely to be encountered in the wilds of \pzip\ codes.
Ideally, an uninformative implicit prior would be used, although it may be complicated to compute from the covariances of the raw data.  
Template fitting codes have an explicit prior input formed by redshifting a number of templates, leading to a highly nonuniform but physically-motivated interim prior.
Machine learning approaches tend to be trained on one of more previously observed data sets that include only galaxies for which spectroscopy is accessible, typically biasing the implicit prior towards atypically bright and/or low redshift populations.
Some efforts have been made to modify an observationally informed implicit prior so that it is more representative of the photometric data for which redshifts are desired \citep{sheldon_photometric_2012}, but, unless it is equal to the true \nz, it will propagate to the results of traditional \nz\ estimation methods.  

\begin{figure*}
	\begin{center}
		\includegraphics[width=0.45\textwidth]{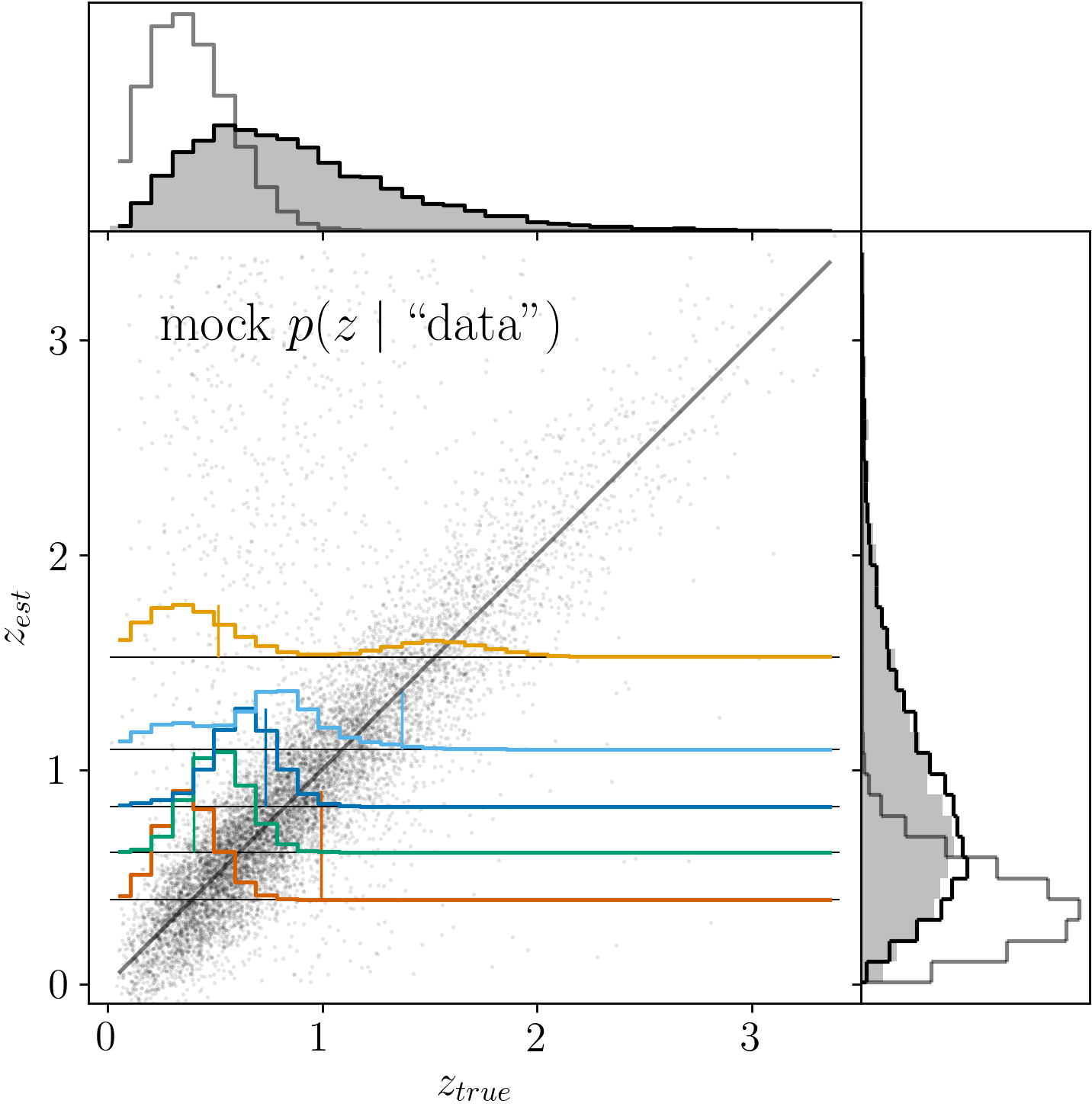}
		\includegraphics[width=0.45\textwidth]{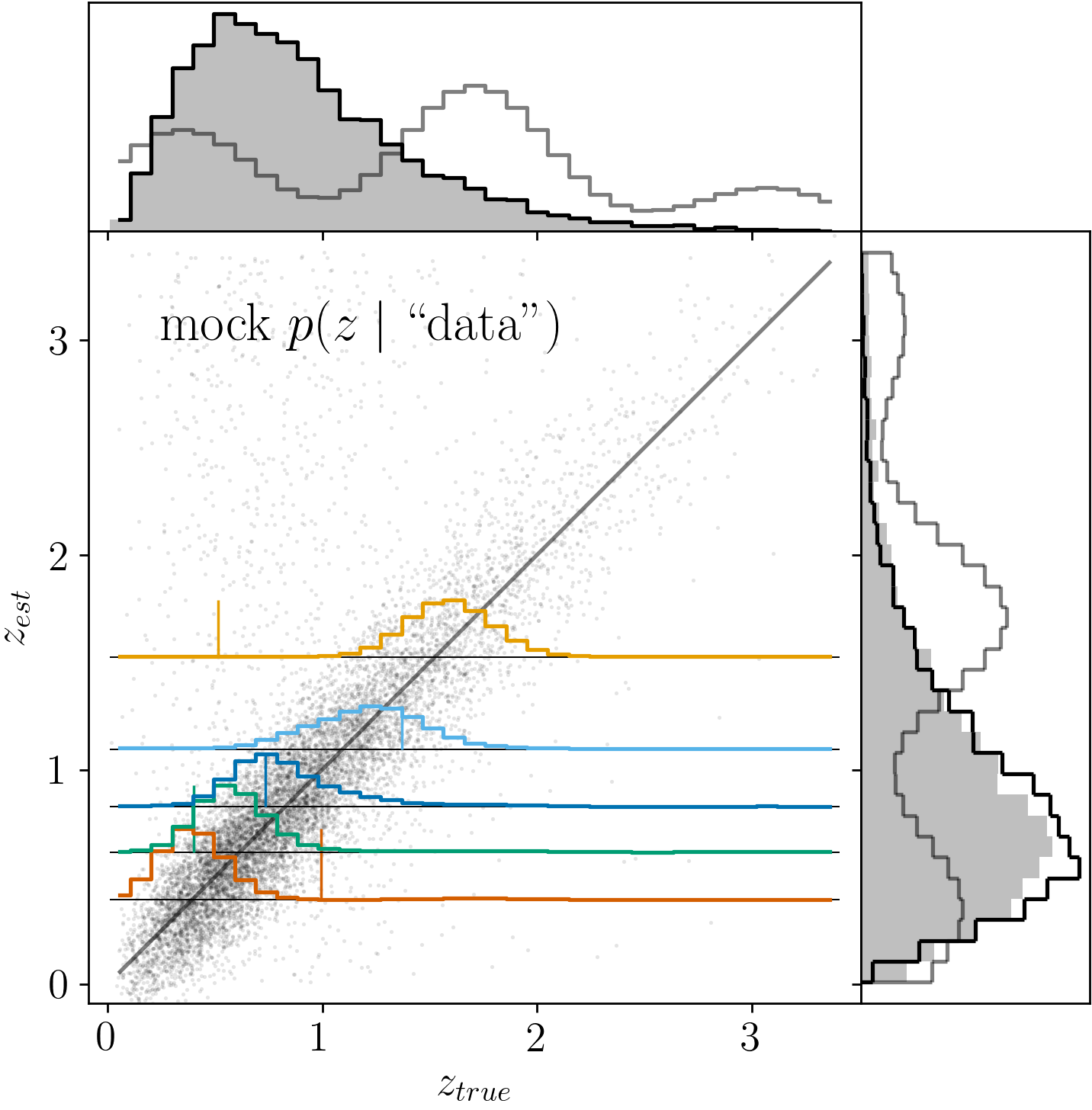}
		\caption{
			Examples of mock \pzip s generated with a machine learning-like implicit prior (left) and a template-fitting-like implicit prior (right), including samples from the probability space of true and observed redshift (black points), \pzip s (colored step functions), the true redshifts of the example \pzip s (colored vertical lines).
			A histogram (light gray) of points in each dimension is shown in the respective inset, with the true redshift distribution (black) and implicit prior (dark gray).
		}
		\label{fig:pzs-priors}
	\end{center}
\end{figure*}

\Fig{fig:pzs-priors} shows examples of \pzip s with a low-redshift favoring implicit prior emulating that of a machine learning approach to \pz\ estimation (left panel) and a more complex interim prior emulating that of a template-fitting \pz\ method (right panel).
One can see that the \pzip s take different shapes from one another even though the marginal histograms of the points are identical.
The machine learning-like implicit prior has been modified to have nonzero value at high-redshift because the implicit prior must be strictly positive definite for the \Chippr\ model to be valid.

\begin{figure*}
	\begin{center}
		\includegraphics[width=0.45\textwidth]{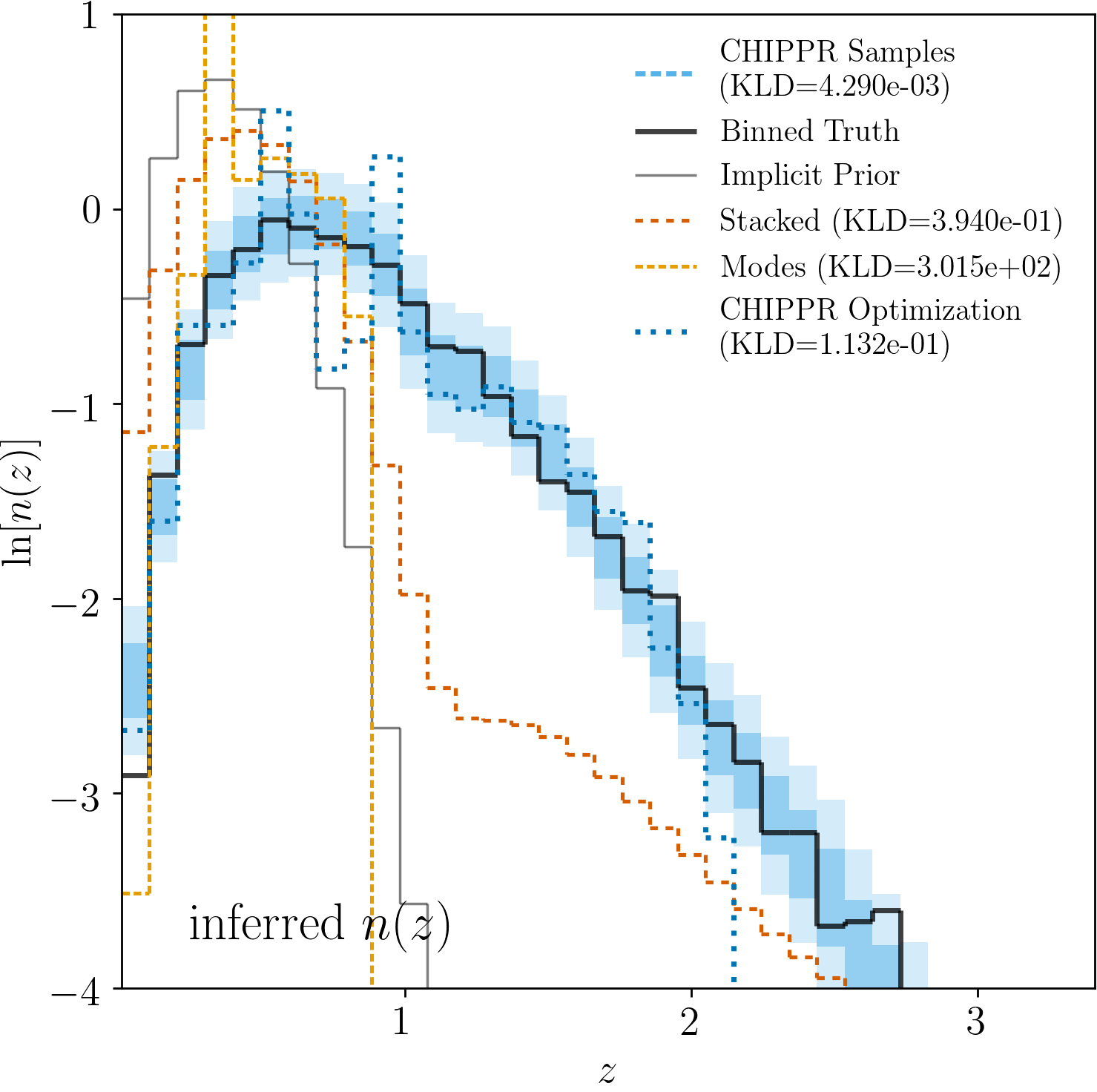}
		\includegraphics[width=0.45\textwidth]{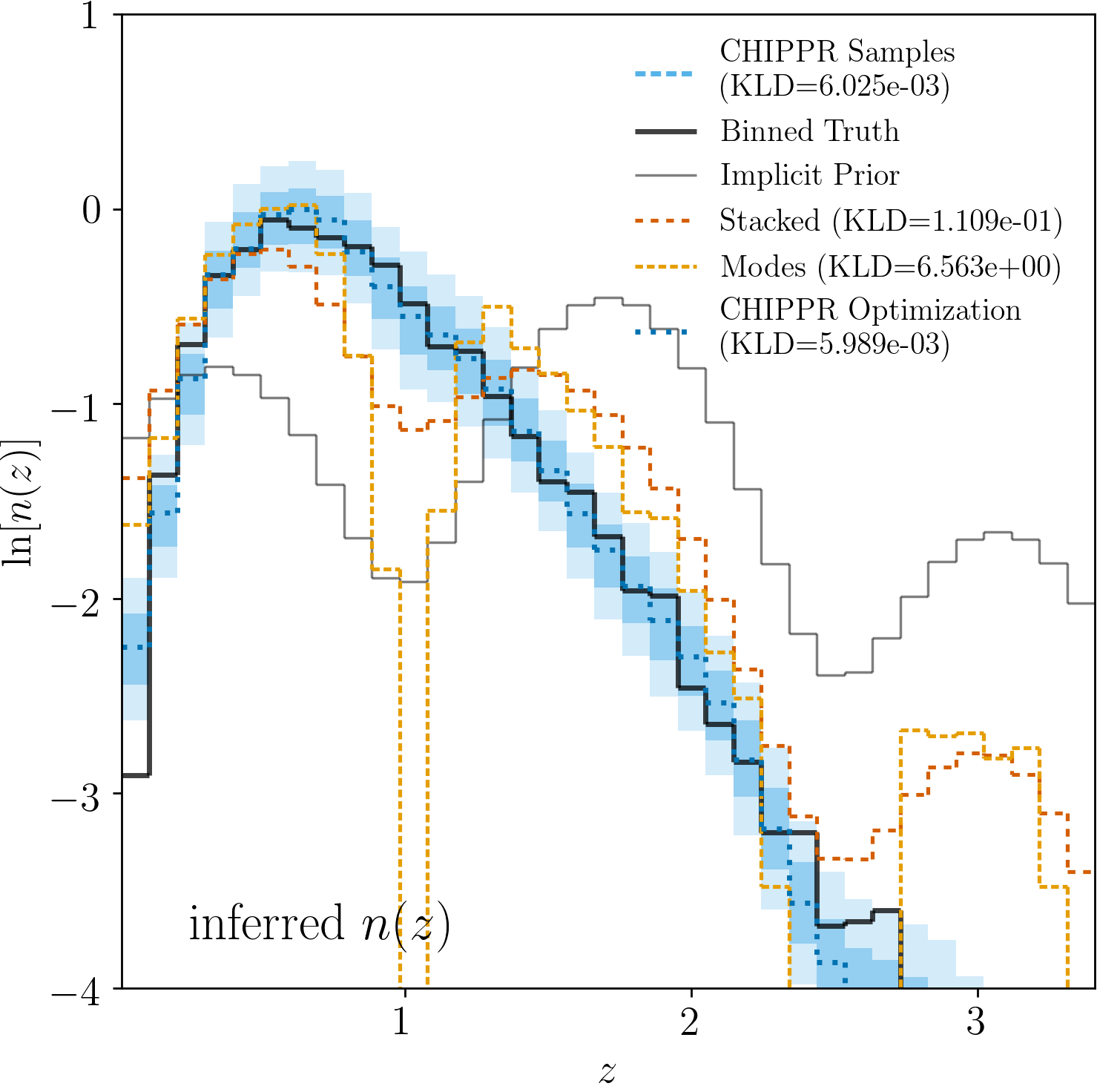}
		\caption
		{The results of \Chippr\ (samples in light blue and optimization in dark blue) and the alternative approaches (the stacked estimator in red and the histogram of modes in yellow) on \pzip s with an implicit prior like that of machine learning \pzip\ approaches (left) and an implicit prior like that of template-fitting \pzip\ codes (right), with the true redshift density (black curve) and implicit prior (gray curve).
			\Chippr\ is robust to a nontrivial implicit prior, but the alternatives are biased toward the implicit prior.
		}
		\label{fig:results-priors}
	\end{center}
\end{figure*}

\Fig{fig:results-priors} shows the performance of \Chippr\ and the traditional methods on \pzip s generated with nontrivial implicit priors.
In both cases, the \Chippr\ \mmle\ effectively recovers the true redshift distribution, and the distribution of \nz\ parameter values reflects higher uncertainty where the implicit prior undergoes large changes in derivative.
The alternatives, on the other hand, are biased by the implicit prior except where it is flat, in the case of high redshifts for the machine learning-like implicit prior, resulting in over $1,000$ times the information loss on \nz\ for the machine learning-like implicit prior and some $5-20$ times the information loss for the template fitting-like implicit prior, relative to the \Chippr\ \mmle.

The main implication of the response of \nz\ estimates to a nontrivial implicit prior is that the implicit prior must be accounted for when using \pzip\ catalogs.

\subsection{Violations of the model}
\label{sec:violations}

\begin{figure*}
	\begin{center}
		\includegraphics[width=0.45\textwidth]{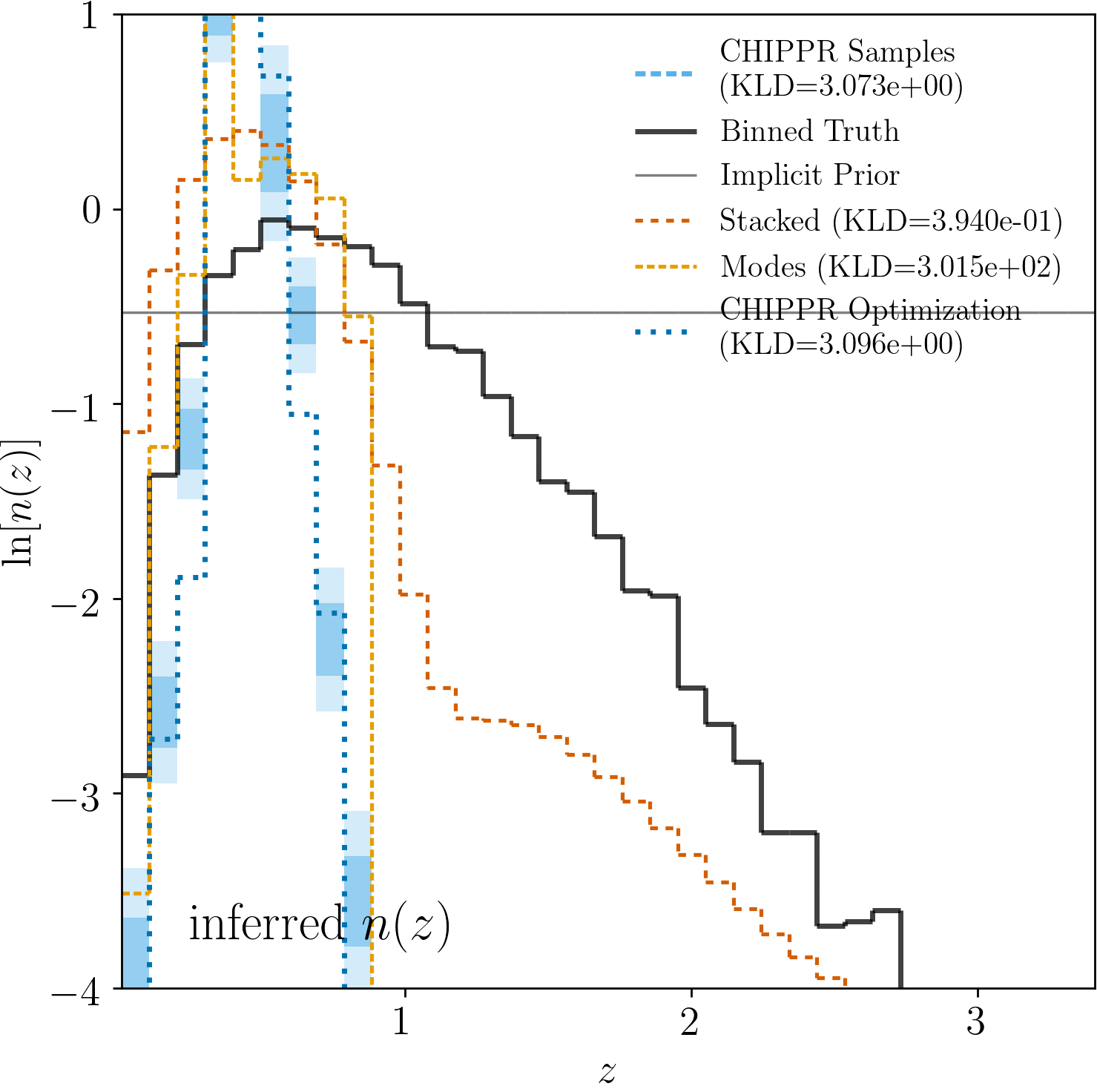}
		\includegraphics[width=0.45\textwidth]{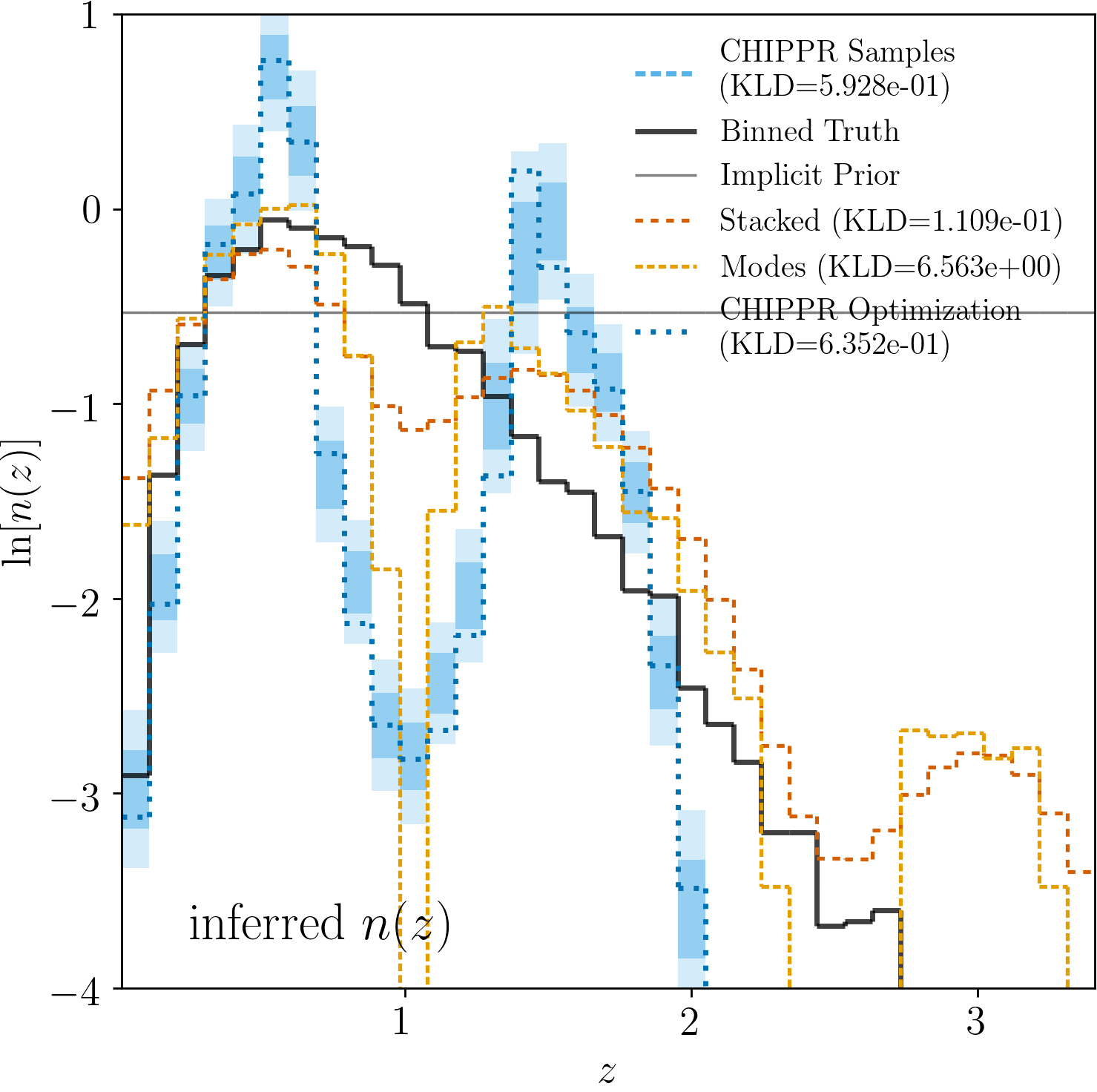}
		\caption
		{The results of \Chippr\ (samples in light blue, optimization in dark blue) and the alternative approaches (the stacked estimator in red, the histogram of modes in yellow) when run with an incorrectly specified implicit prior (gray curve).
			The data upon which each panel's results are based are provided in Figure~\ref{fig:pzs-priors}, where the left corresponds to the sort of implicit prior anticipated of machine learning approaches and the right corresponds to an implicit prior like that of a template-fitting code.
			Here, \Chippr\ has been provided with a uniform implicit prior rather than those used to produce the mock \pzip s, and its performance is notably worse than when it is provided an accurate implicit prior, as in Figure~\ref{fig:results-priors}.
			When the incorrect implicit prior is provided to \chippr, even Bayesian inference cannot recover the true \nz.
		}
		\label{fig:mischaracterized}
	\end{center}
\end{figure*}

In this test, the \pzip s are made to the \lsst\ requirements but the implicit prior used for the inference is not the same as the implicit prior used for generating the data.
\Pzpdf\ codes do not generally provide their implicit prior, with the exception of some template-fitting techniques for which it is a known input.
If we naively used the \pzip\ catalog produced by a generic machine learning or template-fitting code and assumed a flat implicit prior, we would observe the contents of \Fig{fig:mischaracterized}.

The results of using a mischaracterized implicit prior are disastrous, causing every estimator, including \Chippr, to be strongly biased.
The stacked estimator and histogram of modes don't make use of the implicit prior so do no worse than when the implicit prior is accurately provided, but \Chippr\ is sensitive to prior misspecification, which violates the model upon which it is based.
It is thus crucial that \pzip\ methods always characterize and provide the implicit prior.

\section{Conclusion}
\label{sec:con}

This study derives and demonstrates a mathematically consistent inference of a one-point statistic, the redshift density function \nz, based on an arbitrary catalog of \pzpdf s.  
The fully Bayesian \Chippr\ model, based in the fundamental laws of probability, begins with a probabilistic graphical model corresponding to equations for the full posterior distribution over the parameters for \nz.  
The \Chippr\ model is implemented in the publicly available \chippr\ code.
The method is implemented in the publicly available \chippr\ code and validated on mock data.

Using a flexible, self-consistent forward model of the relationship between true and estimated redshifts, capable of encapsulating the complexity of observed redshift-photometry relations (e.g. \Fig{fig:pedagogical_scatter}), we emulate the canonical \pz\ error statistics, intrinsic scatter (\Sect{sec:scatter}), catastrophic outliers (\Sect{sec:outliers}), and canonical bias (\Sect{sec:bias}) one at a time.
Though these test cases may appear overly simplistic, they enable rigorous quantification of the relative performance of each \nz\ estimation techniques under the controlled conditions of each type of error in isolation, at levels equal to and beyond those of \lsst.

Based on our tests, the following statements about the \Chippr\ methodology may be made with confidence:
\begin{itemize}
\item \Chippr\ outperforms traditional estimators of \nz\ under realistically complex conditions, even at pessimistic levels relative to future survey requirements on the traditional \pz\ error statistics, as demonstrated both by eye and according to KLD values corresponding to $10\%$ the information loss of alternative methods.
\item Both the \Chippr\ \mmle\ and the mean of \chippr\ samples are good point estimators of \nz, whereas the histogram of modes is very sensitive to outliers and the stacked estimator is always excessively broad.
\item The error bars on the posterior distribution over \nz\ hyperparameters are interpretable and arise naturally under \Chippr, unlike those that may be assumed for the conventional point estimators.
\end{itemize}
Not only is \Chippr\ the only mathematically correct approach to the problem, it also recovers the true values of the hyperparameters defining \nz\ better than popular alternatives, as measured by the loss of information in \nz.
However, the mathematically valid approach to inference with probabilistic data products incurs nontrivial computational expense, motivating future work to optimize the implementation.

Additionally, this work highlights a crucial and almost entirely overlooked complication to the usage of \pzpdf s, namely the implicit prior, motivating the following recommendations: 
\begin{itemize}
\item In the presence of a nontrivial implicit prior corresponding to the specifics of the architecture of the method by which \pzpdf s are obtained, established methods cannot recover \nz;
a principled hierarchical inference such as \Chippr\ is the only way to recover \nz\ from \pzpdf s.
\item Neither \Chippr\ nor traditional alternatives can recover \nz\ in the presence of a misspecified implicit prior;
the implicit prior used to produce the \pzpdf\ catalog must be known and provided to \Chippr\ in order to recover the true \nz.
\end{itemize}
Given the significance of the implicit prior \citep{schmidt_evaluation_2020}, it is therefore imperative that those developing codes to obtain \pzpdf s provide a way to isolate the implicit prior and that those publishing \pzpdf\ catalogs provide the implicit prior to users.
This mandate is easier said than done, both for template fitting and machine learning approaches.

While the implicit prior is often an explicit input to model-based routines, it may be defined in a space of redshift and SED templates.
In this case, it may not be possible to apply \Chippr\ without marginalizing over additional variables $\psi$ for the SEDs.
In other words, obtaining the implicit prior from a template fitting code may be challenging or even require consideration of higher-dimensional PDFs such as $\pr{z, \mathrm{SED} \gvn \psi^{*}}$.

The situation is more dire for data-driven techniques, whose training sets may not straightforwardly translate into a recoverable implicit prior.
For example, some training set galaxies may contribute to the \pzpdf s more than others, resulting in different effective weights when factoring into, say, a histogram of training set redshifts as the implicit prior.
Additionally, the weights may be stochastic, depending on the random seed used to initialize non-deterministic methods, precluding reproducibility.
It is thus unclear whether the implicit prior can be meaningfully obtained from such methods at all.

A thorough investigation of the degree to which the implicit prior can be meaningfully obtained is outside this paper but should be a priority for all consumers of \pzpdf s.
Alternatively, the trouble with the implicit prior would be avoided altogether if likelihoods were produced rather than posteriors.
We thus encourage the community of those making \pzpdf s to consider developing methods yielding likelihoods rather than posteriors so that the resulting data products may be correctly used in scientific inference more generically.

By showing that \Chippr\ is effective in recovering the true redshift distribution function and posterior distributions on its parameters from catalogs of \pzpdf s, this work supports the production of \pzpdf s by upcoming photometric surveys such as \lsst\ to enable more accurate inference of the cosmological parameters.  
We discourage researchers from co-adding \pzpdf s or converting them into point estimates of redshift and instead recommend the use of Bayesian probability to guide the usage of \pzpdf s.  
We emphasize to those who produce \pzpdf s from data that it is essential to release the implicit prior used in generating this data product in order for any valid inference to be conducted by consumers of this information.
Methodologies for obtaining \pzpdf s must therefore be designed such that there is a known implicit prior, i.e. one that is not implicit at all, so that likelihoods may be recovered.

The technique herein developed is applicable with minimal modification to other one-point statistics of redshift to which we will apply this method in the future, such as the redshift-dependent luminosity function and weak lensing mean distance ratio.  
Future work will also include the extension of this fully probabilistic approach to higher-order statistics of redshift such as the two-point correlation function.

\begin{acknowledgements}
	AIM acknowledges support from the Max Planck Society and the Alexander von Humboldt Foundation in the framework of the Max Planck-Humboldt Research Award endowed by the Federal Ministry of Education and Research.
	During the completion of this work, AIM was supported by National Science Foundation grant AST-1517237 and the U.S. Department of Energy, Office of Science, Office of Workforce Development for Teachers and Scientists, Office of Science Graduate Student Research (SCGSR) program, administered by the Oak Ridge Institute for Science and Education for the DOE under contract number DE‐SC0014664.
	The authors thank Phil Marshall for advice on relevant examples, Elisabeth Krause for assistance with the \cosmolike\ code, Mohammadjavad Vakili for statistical insights, Geoffrey Ryan for programming advice, and Boris Leistedt for other helpful comments in the development of \Chippr.
	This work was completed with generous nutritional support from the Center for Computational Astrophysics.
\end{acknowledgements}


\appendix
\numberwithin{equation}{section}

\section{Derivation}
\label{app:math}

We perform the derivation of \Eq{eqn:fullpost} using log-probabilities.  
What we wish to estimate is then the full log-posterior probability distribution (hereafter the full log-posterior) of the hyperparameters $\ndphi$ given the catalog of photometry $\{\data_{j}\}$.

By Bayes' Rule, the full log-posterior
\begin{equation}
\label{eqn:basicbayes}
\ln[\pr{\ndphi \gvn \{\data_{j}\}}] = \ln[\pr{\{\data_{j}\} \gvn \ndphi}] + \ln[\pr{\ndphi}] - \ln[\pr{\{\data_{j}\}}]
\end{equation}
may be expressed in terms of the full log-likelihood probability distribution (hereafter the full log-likelihood) $\ln[\pr{\{\data_{j}\} \gvn \ndphi}]$ by way of a hyperprior log-probability distribution (hereafter the hyperprior) $\ln[\pr{\ndphi}]$ over the hyperparameters and the log-evidence probability of the data $\ln[\pr{\{\data_{j}\}}]$.
However, the evidence is rarely known, so we probe the full log-posterior modulo an unknown constant of proportionality.

The full log-likelihood may be expanded in terms of a marginalization over the redshifts as parameters, as in 
\begin{equation}
\label{eqn:marginalize}
\ln[\pr{\{\data_{j}\} \gvn \ndphi}] = \ln\left[\integral{\pr{\{\data_{j}\} \gvn \{z_{j}\}} \pr{\{z_{j}\} \gvn \ndphi}}{\{z_{j}\}}\right].
\end{equation}

We shall make two assumptions of independence in order to make the problem tractable; their limitations are be discussed below.  
First, we take $\ln[\pr{\{\data_{j}\} \gvn \{z_{j}\}}]$ to be the sum of $J$ individual log-likelihood distribution functions $\ln[\pr{\data_{j} \gvn z_{j}}]$, as in 
\begin{equation}
\label{eqn:indiedat}
\ln[\pr{\{\data_{j}\} \gvn \{z_{j}\}}] = \sum_{j=1}^{J}\ \ln[\pr{\data_{j} \gvn z_{j}}],
\end{equation}
a result of the definition of probabilistic independence encoded by the box in \Fig{fig:pgm}.
Second, we shall assume the true redshifts $\{z_{j}\}$ are $J$ independent draws from the true $\pr{z \gvn \ndphi}$.  
Additionally, $J$ itself is a Poisson random variable.  
The combination of these assumptions is given by 
\begin{equation}
\label{eqn:indie}
\ln[\pr{\{z_{j}\} \gvn \ndphi}] = -\integral{f(z; \ndphi)}{z} + \sum_{j=1}^{J}\ \ln[\pr{z_{j} \gvn \ndphi}].
\end{equation}
The derivation differs when $J$ is not known, say, when we want to learn about a distribution in nature rather than a distribution specific to data in hand, but for a photometric galaxy catalog where the desired quantity is $n(z)$ for the galaxies entering a larger cosmology calculation, it is a fixed quantity.
A detailed discussion of this matter may be found in \citet{foreman-mackey_exoplanet_2014}.  
Applying Bayes' Rule, we may combine terms to obtain 
\begin{align}
\begin{split}
\label{eqn:posterior}
\ln[\pr{\ndphi \gvn \{\data_{j}\}}] & \propto \ln[\pr{\ndphi}] - \integral{f(z; \ndphi)}{z} + \sum_{j=1}^{J}\ln\left[\integral{\pr{\data_{j} \gvn z} \pr{z \gvn \ndphi}}{z}\right].
\end{split}
\end{align}

Since we only have access to \pzip s, we must be able to write the full log-posterior in terms of log \pzip s rather than the log-likelihoods of \Eq{eqn:posterior}.
To do so, we will need an explicit statement of this implicit prior $\ndphi^{*}$ for whatever method is chosen to produce the \pzip s.  

To perform the necessary transformation from likelihoods to posteriors, we follow the reasoning of \citet{foreman-mackey_exoplanet_2014}.  
Let us consider the probability of the parameters conditioned on the data and an interim prior and rewrite the problematic likelihood of \Eq{eqn:posterior} as 
\begin{align}
\label{eqn:trick}
\begin{split}
\ln[\pr{\data_{j} \gvn z}] = & \ln[\pr{\data_{j} \gvn z}] + \ln[\pr{z \gvn \data_{j}, \ndphi^{*}}] - \ln[\pr{z \gvn \data_{j}, \ndphi^{*}}].
\end{split}
\end{align}

Once the implicit prior $\ndphi^{*}$ is explicitly introduced, we may expand the last term in \Eq{eqn:trick} according to Bayes' Rule to get 
\begin{align}
\begin{split}
\label{eqn:expand}
\ln[\pr{\data_{j} \gvn z}] = & \ln[\pr{\data_{j} \gvn z}] + \ln[\pr{z \gvn \data_{j}, \ndphi^{*}}] + \ln[\pr{\data_{j} \gvn \ndphi^{*}}] - \ln[\pr{z \gvn \ndphi^{*}}] - \ln[\pr{\data_{j} \gvn z, \ndphi^{*}}].
\end{split}
\end{align}
Because there is no direct dependence of the data upon the hyperparameters, we may again expand the term $\ln[\pr{\data_{j} \gvn z, \ndphi^{*}}]$ to obtain 
\begin{align}
\begin{split}
\label{eqn:indterm}
\ln[\pr{\vec{d}_{j} \gvn z}] = & \ln[\pr{\data_{j} \gvn z}] + \ln[\pr{z \gvn \data_{j}, \ndphi^{*}}] + \ln[\pr{\data_{j} \gvn \ndphi^{*}}] - \ln[\pr{z \gvn \ndphi^{*}}]- \ln[\pr{\data_{j} \gvn \ndphi^{*}}] - \ln[\pr{\data_{j} \gvn z}] .
\end{split}
\end{align}
Canceling the undesirable terms for the inaccessible likelihood $\ln[\pr{\data_{j} \gvn z}]$ and trivial $\ln[\pr{\data_{j} \gvn \ndphi^{*}}]$ yields
\begin{equation}
\label{eqn:cancel}
\ln[\pr{\data_{j} \gvn z}] = \ln[\pr{z \gvn \data_{j}, \ndphi^{*}}]  - \ln[\pr{z \gvn \ndphi^{*}}].
\end{equation}
We put this all together to get the full log-posterior probability distribution of 
\begin{align}
\begin{split}
\label{eqn:final}
\ln[\pr{\ndphi \gvn \{\data_{j}\}}] \propto & \ln[\pr{\ndphi}] + \ln \left[\integral{\exp \left[\sum_{j=1}^{J} \left(\ln[\pr{z \gvn \data_{j}, \ndphi^{*}}] + \ln[\pr{z \gvn \ndphi}] - \ln[\pr{z \gvn \ndphi^{*}}] \right)\right]}{z}\right] ,
\end{split}
\end{align}
which is equivalent to that of \citet{hogg_inferring_2010}, though the context differs.

The argument of the integral in the log-posterior of \Eq{eqn:final} depends solely on knowable quantities (and those we must explicitly assume) and can be calculated for a given sample of log \pzip s $\{\ln[\pr{z \gvn \data_{j}, \ndphi^{*}}]\}$ and the implicit prior $\pr{z \gvn \ndphi^{*}}$ with which they were obtained, noting the relation of 
\begin{equation}
\label{eqn:params}
\pr{z \gvn \ndphi} = \frac{f(z; \ndphi)}{\integral{f(z; \ndphi)}{z}}.
\end{equation}
Since we cannot know constant of proportionality, we sample the desired full log-posterior $\ln[\pr{\ndphi \gvn \{\data_{j}\}}]$ using Monte Carlo-Markov chain (MCMC) methods.

\bibliographystyle{apj}
\bibliography{ms}

\end{document}